\documentclass[aps,prd,twocolumn,superscriptaddress,nofootinbib]{revtex4}
\usepackage{hyperref}
\usepackage{amsmath,amssymb,amsfonts}
\usepackage{color}
\usepackage{graphicx}
\usepackage{enumerate} 
\usepackage{colordvi} 
\usepackage{bm}
\usepackage{multirow}
\usepackage{braket}
\usepackage{dcolumn}

\newcommand{\ipmu}{Kavli IPMU (WPI), UTIAS, The University of Tokyo, 5-1-5 Kashiwanoha, Kashiwa, Chiba 277-8583, Japan}
\newcommand{\ntu}{Department of Physics, National Taiwan University, No. 1, Section 4, Roosevelt Road, Taipei 106216, Taiwan}
\newcommand{\asiaa}{Academia Sinica Institute of Astronomy and Astrophysics (ASIAA), No. 1, Section 4, Roosevelt Road, Taipei 106216, Taiwan}
\newcommand{\yitp}{Center for Gravitational Physics and Quantum Information, Yukawa Institute for Theoretical Physics, Kyoto University, Kyoto 606-8502, Japan}
\newcommand{\kmi}{Kobayashi-Maskawa Institute for the Origin of Particles and the Universe, 
Nagoya University, Nagoya, 464-8602, Japan}
\newcommand{\nagoya}{Institute for Advanced Research, Nagoya University, Furo-cho Chikusa-ku, Nagoya 464-8601, Japan}

\begin{document}

\title{Testing local position invariance with odd multipoles of galaxy clustering statistics}

\author{Takuya Inoue}
\email{takuya.inoue@yukawa.kyoto-u.ac.jp}
\affiliation{\ntu}
\affiliation{\asiaa}
\affiliation{\yitp}

\author{Teppei Okumura}
\affiliation{\asiaa}
\affiliation{\ipmu}

\author{Shohei Saga}
\affiliation{\kmi}
\affiliation{\nagoya}

\author{Atsushi Taruya}
\affiliation{\yitp}
\affiliation{\ipmu}

\date{\today}

\begin{abstract}
We investigate cosmological constraints on local position invariance (LPI), a key aspect of the Einstein equivalence principle (EEP), through asymmetric galaxy clustering. 
The LPI asserts that the outcomes of the nongravitational experiments are identical regardless of location in spacetime and has been tested through measurements of the gravitational redshift effect.
Therefore, measuring the gravitational redshift effect encoded in galaxy clustering provides a powerful and novel cosmological probe of the LPI.
Recent work by Saga {\it et al.} proposed its validation using the cross-correlation function between distinct galaxy samples, but their analysis focused solely on the dipole moment. In this paper, we extend their work by further analyzing a higher-order odd multipole moment, the octupole moment, in the constraints on the LPI-violating parameter, $\alpha$, expected from galaxy surveys such as Dark Energy Spectroscopic Instrument, Euclid space telescope, Subaru Prime Focus Spectrograph, and Square Kilometre Array. We demonstrate that combining the octupole and dipole moments significantly improves the constraints, particularly when the analysis is restricted to larger scales, characterized by a large minimum separation $s_{\rm min}$. For a conservative setup with $s_{\rm min}=15 {\rm Mpc}/h$, we find an average improvement of 11$\%$ compared to using the dipole moment alone. Our results highlight the importance of higher-order multipoles in constraining $\alpha$, providing a more robust approach to testing the EEP on cosmological scales.

\end{abstract}

\maketitle
\section*{Introduction}
The origin of the current accelerated expansion of the Universe remains unknown, making it essential to test gravity on cosmological scales. Galaxy surveys provide a powerful tool for this purpose, with two-point statistics of galaxy distribution widely employed. 
The standard Doppler effect caused by galaxy peculiar velocities, known as redshift-space distortion (RSD), introduces an apparent anisotropy in the galaxy distribution along the line of sight. 
This anisotropy, related to the growth rate of structure formation $f\equiv d{\rm ln} D_+(a)/d{\rm ln}a$, where $D_+(a)$ is the linear growth factor as a function of the scale factor $a$, has been measured using even multipole moments of galaxy correlation functions, offering a means to probe gravity \cite{Sargent_1977, Peebles_1980, Kaiser_1987, Linder_2008, Guzzo_2008, Alam_2017}.

Beyond the standard Doppler effect, weaker relativistic effects can provide a new approach to testing gravity on cosmological scales \cite{Sasaki_1987,Matsubara_2000,Yoo_2009,Bonvin_2011,Challinor_2011,Yoo_2012,Jeong_2012,Yoo_2014,Raccanelli_2016}. These effects contribute to asymmetric (odd) multipole moments in cross-correlations between different galaxy populations \cite{McDonald_2009, Croft_2013, Bonvin_2014_1,Bonvin_2014,Kodwani_2019}. Importantly, odd multipole moments have been proposed as probes of the weak equivalence principle, a fundamental pillar of the Einstein equivalence principle (EEP), by detecting potential differences in the motion of baryons and dark matter on large scales where the Euler equation governs the motion of matter \cite{Bonvin_2018, Bonvin_2020, Umeh_2021}. This makes relativistic effects a more fundamental test of gravity compared to conventional RSD measurements of $f$.

Among these relativistic effects, the gravitational redshift has been observed in galaxy clusters \cite{Wojtak_2011, Sadeh_2015, Jimeno_2015} and in galaxy clustering on nonlinear scales ($<10\ \rm Mpc/h$) \cite{Alam_2017_gr}, but its detection on larger scales ($>20\ \rm Mpc/h$) remains elusive \cite{Gaztanaga_2017, Alam_2017_gr}. On small scales, significant deviations from linear theory have been reported in the dipole moments of cross-correlations measured from $N$-body simulations, where gravitational redshift dominates \cite{Zhu_2017, Breton_2018, Beutler_2020}. 
References~\cite{Saga_2020, Saga_2022} developed an analytic model incorporating nonlinear halo potentials, successfully reproducing the results of Ref.~\cite{Breton_2018} (see also Refs.~\cite{Giusarma_2017,  Dio_2019} for predictions beyond linear theory). Using this model, Ref.~\cite{Saga_2022} predicted significant detections of the dipole in ongoing and upcoming galaxy surveys, such as the Subaru Prime Focus Spectrograph (PFS) \cite{PFS2014}, the Dark Energy Spectroscopic Instrument (DESI) \cite{DESI2016}, the Euclid space telescope \cite{Euclid_2024}, and the Square Kilometre Array (SKA) \cite{Bacon_2020} (see also Refs.~\cite{Bonvin_2023} and \cite{Lepori_2024} for the detectability of the relativistic dipole for DESI and Euclid, respectively). 
Reference~\cite{Saga_2023} further demonstrated that such detections offer a cosmological test of local position invariance (LPI), another fundamental pillar of the EEP stating that the physical laws are independent of the spacetime location where the experiment is performed, with precision comparable to that of ground-based experiments~\cite{Pound_1959, Pound_1965}.

In this Letter, we focus for the first time on the octupole moment of the cross-correlation function between two galaxy populations on small scales, induced by the interplay between nonlinear halo potential and standard Doppler contributions. Using Fisher analysis, we assess its impact on constraining the LPI-violating parameter $\alpha$, which quantifies deviations from the gravitational redshift $z_{\rm grav}$ induced by the potential difference $\Delta \phi$ under EEP \cite{Will_2018}, $z_{\rm grav}=(1+\alpha) \Delta \phi$. 
Such deviations may arise in models where the electromagnetic sector couples to a scalar field, leading to spacetime variations in the fine-structure constant \cite{Bekenstein_1982, Sandvik_2002, Barrow_1998, Barrow_1999}. While general relativity predicts that the gravitational redshift depends only on $\Delta \phi$, $\alpha$ encapsulates these additional spacetime variations.
We further show how combining the octupole with the dipole improves constraints on $\alpha$, accounting for the dependence on the minimum separation between samples in the cross-correlation\footnote{Recently, Ref.~\cite{Blanco_2024} also investigated the benefit of utilizing the octupole moment but on large scales induced by the wide-angle effect as a probe of the magnification bias, evolution bias, and cosmological models.}.

Throughout this paper, we assume a flat $\Lambda {\rm CDM}$ model with fiducial cosmological parameters from the 7 year WMAP results \cite{Komatsu:2011}.

\section*{Asymmetric correlation function}
We present the analytic model of the cross-correlation function between two galaxy populations, derived in Ref.~\cite{Saga_2022}.
The galaxy distribution in redshift surveys appears distorted not only by the standard Doppler effect but also by various relativistic effects (e.g., Refs.~\cite{Yoo_2009, Bonvin_2011, Challinor_2011}). When considering only dominant contributions at small scales, the galaxy positions in redshift space, $\bm{s} $, are related to those in real space, $\bm{x}$, by \cite{Saga_2020, Saga_2022}
\begin{align}
\bm{s} =  \bm{x}+\frac{1}{aH}\left(\bm{v} \cdot \bm{\hat{x}}\right)\bm{\hat{x}}+\epsilon_{\rm NL}\bm{\hat{x}}
\label{eq:map}~,
\end{align}
where the second term represents the standard Doppler effect and $H$ and $\bm{v}$ are the Hubble parameter, and the peculiar velocity of galaxy, respectively.
The hat denotes the unit vector. 
The third term represents the nonlinear effects within host halos as nonperturbative contributions with $\epsilon_{\rm NL}$ given as \cite{Saga_2023}
\begin{align}
\epsilon_{\rm NL}=-\frac{1}{aH}\left(\phi_{\rm halo}+\frac{5}{2}v^2_{\rm g}\right)
\label{eq:enl}~,
\end{align}
where the first term is the gravitational redshift effect caused by the nonlinear halo potential at the galaxy position, $\phi_{\rm halo}$.
The second term, $v^2_{\rm g}$, is the velocity dispersion, consisting of two contributions, $v^2_{\rm g}=v^2_{\rm vir}+v^2_{\rm halo}$, with $v^2_{\rm vir}$ being the virial motion of galaxies within halos and $v^2_{\rm halo}$ being the coherent motion of halos on large scales. The latter further contains contributions of the transverse Doppler, light-cone, and surface brightness modulation effects \cite{Zhao_2013,  Kaiser_2013,Cai_2017,Zhu_2017,Breton_2018}. 
Following Refs.~\cite{Saga_2020,Saga_2022,Saga_2023}, we model $\phi_{\rm halo}$ and $v^2_{\rm vir}$ using the Navarro-Frenk-White density profile, where they are obtained from the Poisson and Jeans equations, respectively \cite{Navarro_1996,Lokas_2001}. The velocity dispersion $v^2_{\rm halo}$ is calculated using linear peak theory for Gaussian density fields \cite{Sheth_2001}. 
Galaxies are not always located at the centers of their host halos but instead reside in shallower regions, offset from the center of the gravitational potential. This causes virialized motions of galaxies and the suppression of the gravitational redshift effect. To account for this off-centering effect, we model galaxy positions within halos as the  Gaussian distribution with the dispersion $R_{\rm off}$. Consequently, both the halo potential $\phi_{\rm halo}$ and the galaxy velocity dispersion $v^2_{\rm g}$ are expressed as functions of the halo mass $M$, redshift $z$, and off-centering parameter $R_{\rm off}$, i.e., $\phi_{\rm halo}(M, z, R_{\rm off})$ and $v^2_{\rm g}(M, z, R_{\rm off})$ \cite{Hikage_2013, Masaki_2013, Yan_2020, Saga_2022, Saga_2023}. In our analysis, we relate $M$ to the linear galaxy bias $b$ via the Sheth-Tormen mass function \cite{Sheth_1999}. $b$ and $R_{\rm off}$ are marginalized over to estimate the uncertainties in the LPI-violating parameter $\alpha$.
Given the mapping in Eq.~(\ref{eq:map}), the galaxy density field in the redshift space, $\delta^{(\rm S)}(\bm{s})$, is obtained as
$\delta^{(\rm S)}(\bm{s}) =  \int{\rm d}^{3}\bm{k}/(2\pi)^{3}{e}^{{i}\bm{k}\cdot\bm{s}}\delta^{(\rm S)}(\bm{k})$,
where the galaxy density field in Fourier space, $\delta^{(\rm S)}(\bm{k})$, is the sum of the standard contribution, $\delta^{(\rm std)}(\bm{k})$, and the additional contribution from the gravitational redshift effect, $\delta^{(\epsilon_{\rm NL})}(\bm{k})$, given by \cite{Saga_2022}
\begin{align}
&
\delta^{(\rm std)}(\bm{k})=\left(b+f\mu_k^2-if\frac{2 \mu_k}{ks}\right)\delta_{\rm m}(\bm{k}) ~, \\ &
\delta^{(\epsilon_{\rm NL})}(\bm{k})=\frac{\epsilon_{\rm NL}}{s}\left(-1+\mu^2_k-if\frac{2}{ks}\mu_k-ibks\mu_k \right. \notag \\ & \qquad \qquad \left. -2f\mu^2_k-i\frac{2}{ks}\mu_k-ifks\mu^3_k \right)\delta_{\rm m}(\bm{k}) \label{eq:delta_enl}~,
\end{align}
where $\delta_{\rm m}(\bm{k})$ is the linear matter density field, and $\mu_k$ is the directional cosine between the wave vector and the line-of-sight direction. 

To extract the asymmetric components from the galaxy density field, we compute the cross-correlation function between two galaxy populations X and Y, $\xi^{\rm XY}(\bm{s_1},\bm{s_2}) \equiv \Braket{\delta^{(\rm S)}_{\rm X}(\bm{s}_{1})\delta^{(\rm S)}_{\rm Y}(\bm{s}_{2})}$ \cite{Bonvin_2014}. 
Accounting for the triangular geometry formed by the two galaxy positions and the observer's position,
the correlation function is characterized by three variables: the separation, $s=|\bm{s}_{2}-\bm{s}_{1}|$; the directional cosine between the separation vector and the line-of-sight vector, $\mu_{s}=\hat{\bm s} \cdot \hat{\bm d}$; and the midpoint line-of-sight distance\footnote{Following Ref.~\cite{Saga_2023}, we adopt this definition for our line-of-sight definition. However, there are other definitions, such as the bisector or the end point vector (see Refs.~\cite{Taruya_2019,Reimberg_2016,Castorina_2018a,Castorina_2018b} for the investigation of the choice of different line-of-sight definitions).} $d=|(\bm{s}_{1}+\bm{s}_{2})/2|$, i.e., $\xi^{\rm XY}(\bm{s_1},\bm{s_2})=\xi^{\rm XY}(s,\mu_s,d)$. 
We separate the dependence on the line-of-sight distance from the correlation function by expanding in powers of ($s/d$): $\xi^{\rm XY}(s,\mu_s,d)=\sum_{n=0}(\frac{s}{d})^n \xi^{\rm XY}_{n}(s,\mu_s)$
where the $n=0$ term corresponds to the expression under the plane-parallel approximation.
We are interested in the asymmetric components of the correlation function, denoted as $\xi^{\rm XY}_{\rm asym}$. It is expressed in terms of odd powers of $\mu_s$:
\newpage
\begin{widetext}
\begin{align}
\xi^{\rm XY}_{\rm asym}(s, \mu_s, d)  
&= 
\frac{\Delta \epsilon_{\rm NL}}{s}  \Biggl[-\Bigl\{b_{\rm X}b_{\rm Y}+\frac{3}{5}\left(b_{\rm X}+b_{\rm Y}\right)f+\frac{3}{7}f^2\Bigr\}\mathcal{L}_{1}(\mu_s)\Xi^{(-1)}_{1}(s)
+\frac{2}{45}f\Bigl\{9\left(b_{\rm X}+b_{\rm Y}\right)+10f\Bigr\}\mathcal{L}_{3}(\mu_s)\Xi^{(-1)}_{3}(s)\notag \\
&
-\frac{8}{63}f^2\mathcal{L}_{5}(\mu_s)\Xi^{(-1)}_{5}(s)\Biggr]+\frac{2}{15}\Delta bf \frac{s}{d} \Biggl[5\mathcal{L}_{1}(\mu_s)\Xi^{(0)}_{0}(s) +\Bigl\{2\mathcal{L}_{1}(\mu_s)+3\mathcal{L}_{3}(\mu_s)\Bigr\}\Xi^{(0)}_{2}(s)\Biggr] 
~,
\label{eq:ximu}
\end{align}
\end{widetext}
where $\Delta \epsilon_{\rm NL}=\epsilon_{\rm NL,X}-\epsilon_{\rm NL,Y}$, $\Delta b=b_{\rm X}-b_{\rm Y}$, and $\mathcal{L}_{\ell}(\mu_s)$ represent the Legendre polynomials. 
The function $\Xi^{(m)}_\ell(s)$ is defined as 
$\Xi^{(m)}_{\ell}({s}) = \int \frac{k^{2}{\rm d}{k}}{2\pi^{2}}\frac{j_{\ell}({ks})}{(ks)^{m}} P_{\rm L}(k)$
where $j_{\ell}(ks)$ and $P_{\rm L}(k)$ are the spherical Bessel function and linear matter power spectrum, respectively.
In Eq.~(\ref{eq:ximu}), we included the contributions up to $n=1$, where the terms proportional to $(s/d)$ represent the leading-order wide-angle correction, while the remaining terms are derived under the plane-parallel approximation.
To characterize the anisotropies, we use the multipole expansion of the correlation function, $\xi^{{\rm XY}}_{\ell}(s,d)= \frac{2 \ell+1}{2} \int^{1}_{-1}d \mu_s \mathcal{L}_{\ell}(\mu_s) \xi^{\rm XY}(s, \mu_s, d)$.
Substituting Eq.~(\ref{eq:ximu}) yields the odd multipole moments (see Appendix C of Ref.~\cite{Saga_2022}).
While the octupole ($\ell=3$) is induced by the interplay between the Doppler and gravitational redshift effects as well as the wide-angle correction of the Doppler effect, the triakontadipole ($\ell=5$) is induced solely by the former.
Throughout this paper, we investigate the impact of the octupole, the next-leading odd multipole, on the constraints on the LPI violation parameter $\alpha$. 
The triakontadipole contribution turns out to be negligible, and we do not consider it here.

\begin{figure*}
\centering
\includegraphics[width=0.64\textwidth]{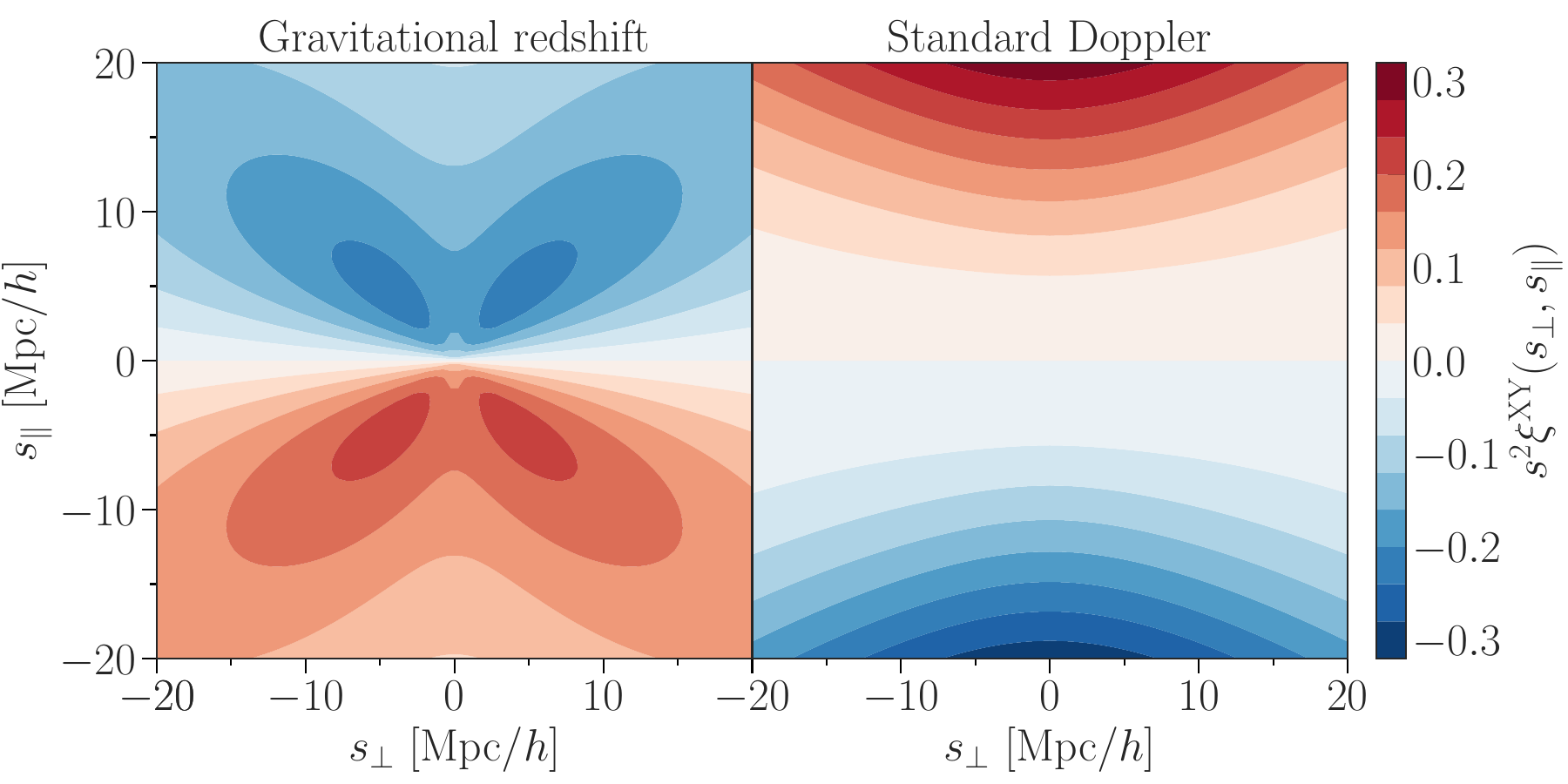}
\hfill
\includegraphics[width=0.35\textwidth]{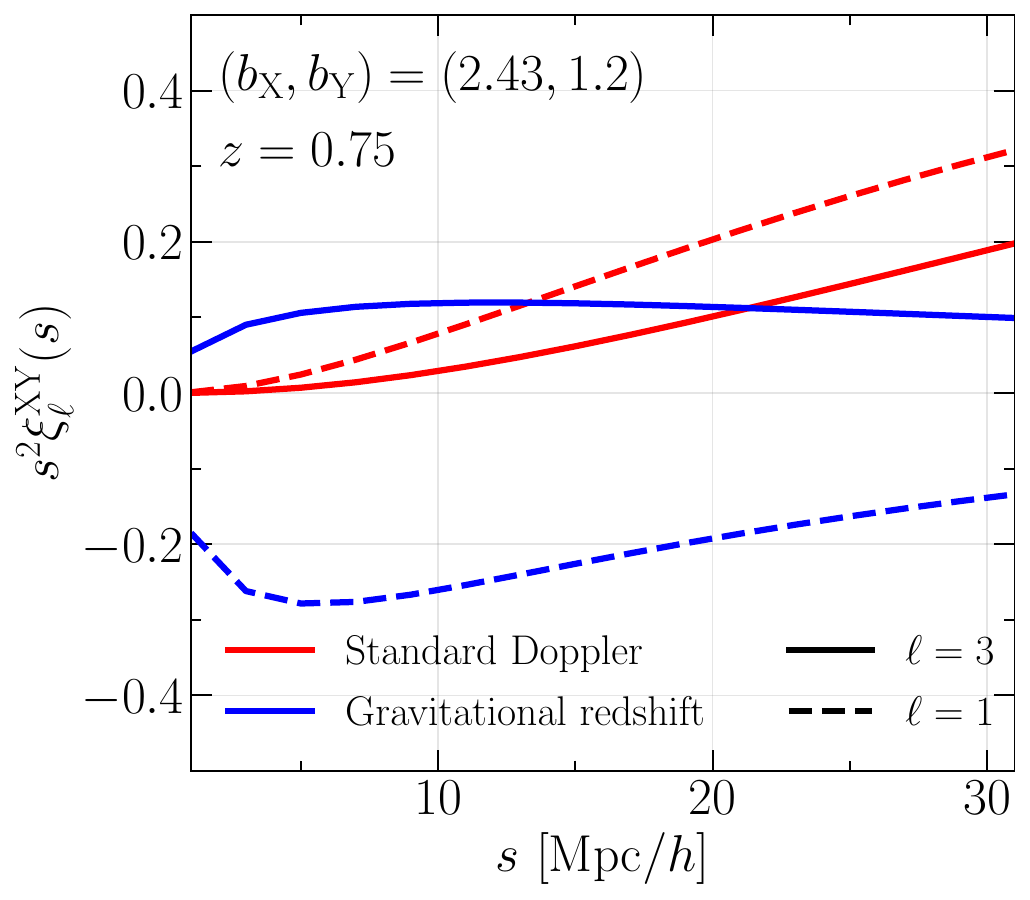}
\caption{{\it Left panels}: two-dimensional cross-correlation function between different populations ($b_{\rm X},b_{\rm Y})=(2.43,1.2$) at $z=0.75$ as a function of separations perpendicular and parallel to the line of sight, $(s_{\perp}, s_{\parallel})=(s\sqrt{1-\mu^2_s}, s\mu_{s})$, in redshift space, $s^2\xi^{\rm XY}(s_{\perp}, s_{\parallel})$. The left and right contours show the additional gravitational redshift and standard Doppler contributions, respectively.
{\it Right panel}: dipole (dashed) and octupole (solid) obtained from the gravitational redshift (blue) and standard Doppler (red) effects.}
\label{fig:xi2D}
\end{figure*}

Before proceeding to the analysis, let us look at behaviors of the asymmetric galaxy clustering on small scales.
The left two panels of Fig.~\ref{fig:xi2D} show the asymmetric components of the full two-dimensional correlation function obtained from the gravitational redshift and standard Doppler contributions, and the right panel shows its dipole and octupole moments. The gravitational redshift and standard Doppler effects are computed from the terms proportional to $\Delta \epsilon_{\rm NL}$ and $\Delta b$ in Eq.~(\ref{eq:ximu}), respectively. 
The results exhibit distinct anisotropic features that cannot be explained by the dipole anisotropy alone. Such features manifest prominently in the octupole, as shown in the right panel.
The standard Doppler effect exhibits positive odd multipoles, while the dipole and octupole induced by the gravitational redshift show negative and positive amplitudes, respectively.
This indicates that there is no sign flip in the octupole, unlike the dipole discussed in Refs.~\cite{Breton_2018, Saga_2020, Saga_2022}.

\section*{Forecast formalism}
Here, we present the Fisher matrix formalism \cite{Tegmark_1997,Tegmark_etal_1997}. We consider five free parameters that characterize the asymmetric components \cite{Saga_2023}: $\bm{\theta} \equiv (\alpha, R_{{\rm off}, {\rm X/Y}}, b_{\rm X/Y})$, where $R_{\rm off}$ and $b$ are nuisance parameters to be marginalized over. We set the fiducial value of $\alpha$ to zero. We fix the fiducial value of $R_{\rm off}$ to $R_{\rm off}=0.2 r_{\rm vir}$ and incorporate the expected errors $\sigma_{R_{\rm off}}=0.01 r_{\rm vir}$ as a Gaussian prior, where $r_{\rm vir}$ is the virial radius of halos \cite{Hikage_2013, Masaki_2013, Yan_2020, Saga_2023}. The fiducial values of $b$ are determined based on the galaxy survey samples (see the next section), while the expected errors, $\sigma_b$, calculated from another Fisher matrix using the even multipoles of galaxy clustering at large scales, are incorporated as Gaussian priors (see Appendix C of Ref.~\cite{Saga_2023} for the detailed calculation). Note that the linear growth rate $f$ is fixed by the WMAP cosmology.

Given the model for the odd multipoles with the free parameters, the Fisher matrix for the $n$th redshift slice $z_{n}$ is computed as: 
\begin{align}
&
F_{n,ij} 
=
\sum^{s_{{\rm max}}}_{s_{1,2}=s_{\rm min}}\sum^{N}_{a,b}\frac{\partial \xi^{\rm XY}_{a}(s_1,z_n)}{\partial \theta_i}[\mathcal{C}^{-1}]_{ab}\frac{\partial  \xi^{\rm XY}_{b}(s_2,z_n)}{\partial \theta_j}
, \label{eq: def_Fisher_multipole}
\end{align} 
where $s_{{\rm min}}$ and $s_{{\rm max}}$ are the minimum and maximum separations, respectively, and $N$ is the number of multipole moments included in the analysis: $N=1$ when we use a single multipole and $N=2$ when we use both the dipole and octupole. 
The data vector of multipoles $\xi^{\rm XY}_{a}(s_1,z_n)$ and the full covariance matrix $\mathcal{C}_{ab}$ in Eq.~(\ref{eq: def_Fisher_multipole}) are, respectively, given as 
$\xi^{\rm XY}_{a}(s_1,z_n)=$ $\{\xi^{\rm XY}_{1}, \ \xi^{\rm XY}_{3}\}$
and
\begin{equation}
\mathcal{C}_{ab}(s_1,s_2,z_n)=
\begin{pmatrix}
{\rm{Cov}^{\rm XY}_{11}} & {\rm{Cov}^{\rm XY}_{13}}\\
{\rm{Cov}^{\rm XY}_{31}} & {\rm{Cov}^{\rm XY}_{33}}
\end{pmatrix}. \label{eq: def_all_cov}
\end{equation}
The covariance matrix, ${\rm Cov}^{\rm XY}_{\ell\ell'}(s_1,s_2,z)$, defined as ${\rm Cov}^{\rm XY}_{\ell\ell'}(s_1,s_2,z) \equiv \Braket{\hat{\xi}^{\rm XY}_{\ell}(s_1,z)\hat{\xi}^{\rm XY}_{\ell'}(s_2,z)}-\Braket{\hat{\xi}^{\rm XY}_{\ell}(s_1,z)}\Braket{\hat{\xi}^{\rm XY}_{\ell'}(s_2,z)}$ where $\hat{\xi}^{\rm XY}_{\ell}$ represents the estimator for the multipole moment, is analytically derived under the plane-parallel approximation and Gaussianity of the observed density field (see Refs.~\cite{Hall_2017, Saga_2022} for the derivation):
\begin{widetext}
\begin{align}
{\rm Cov}^{\rm XY}_{\ell\ell'}(s_1,s_2,z)
& 
= \frac{(2\ell+1)(2\ell'+1)}{V}\int\frac{k^{2}{\rm d}k}{2\pi^{2}}\; j_{\ell}(ks_1)j_{\ell'}(ks_2)i^{\ell-\ell'}
\sum_{\ell_{1},\ell_{2}}G^{\ell_{2}\ell_{1}}_{\ell'\ell}\Biggl[ P^{\rm XX}_{\ell_{1}}P^{\rm YY}_{\ell_{2}} +(-1)^{\ell'} P^{\rm XY}_{\ell_{1}}P^{\rm XY}_{\ell_{2}}\Biggr]
\notag \\
& 
+ \frac{(2 \ell+1)(2 \ell'+1)}{V}\int\frac{k^{2}{\rm d}k}{2\pi^{2}}\; j_{\ell}(ks_1)j_{\ell'}(ks_2)i^{\ell-\ell'} 
\sum_{L}
\left(\begin{array}{c c c}
L & \ell & \ell'\\
0 & 0 & 0
\end{array}
\right)^{2}
\Biggl[
\frac{P^{\rm XX}_{L}}{n_{\rm Y}} + \frac{P^{\rm YY}_{L}}{n_{\rm X}}
\Biggr]
\notag \\
&
+ \frac{\delta^{\rm K}_{s,s'}\delta^{\rm K}_{\ell,\ell'}}{4\pi s^{2} L_{\rm p}}\frac{2\ell+1}{n_{\rm X}n_{\rm Y}V} ~,
\label{eq:multipole covariance}
\end{align}
\end{widetext}
where $V$, $n_{\rm X/Y}$, and $L_{\rm p}$ are the survey volume, mean galaxy number density, and side-length of square pixels, respectively.
The coefficient $G^{\ell_{2}\ell_{1}}_{\ell'\ell}$ is expressed in terms of the Wigner 3-j symbols as
\begin{align}
G^{\ell_{2}\ell_{1}}_{\ell'\ell}
= \sum_{\ell_{3}}(2\ell_{3}+1)
\left(
\begin{array}{c c c}
\ell_{1} & \ell_{2} & \ell_{3}\\
0 & 0 & 0
\end{array}
\right)^{2}
\left(
\begin{array}{c c c}
\ell & \ell' & \ell_{3}\\
0 & 0 & 0
\end{array}
\right)^{2}
~ . \label{eq: Gell}
\end{align}
The function $P^{\rm XY}_{\ell}$ is the Fourier counterpart of the correlation function multipole under the plane-parallel approximation, 
$\xi^{\rm XY}_{\ell}(s,z) = (-{\rm i})^{\ell}\int \frac{k^{2}\, {\rm d}k}{2\pi^{2}}\,
j_{\ell}(ks)P^{\rm XY}_{\ell}(k,z)$. 

It is worth noting that the plane-parallel approximation remains valid for the covariance matrix on scales below $190 {\rm Mpc}/h$ \cite{Lepori_2018}.

For the $n$th redshift bin in galaxy surveys, the one-dimensional error on the LPI-violating parameter $\alpha$ is obtained from the inverse Fisher matrix as $\sigma_{n, \alpha}=\sqrt{\left(F_n^{-1} \right)_{\alpha \alpha}}$ after the nuisance parameters are marginalized over. When combining all the redshift bins, the expected one-dimensional error on $\alpha$ for a given survey is given by $\sigma_{\alpha}=1/\sqrt{\sum_{n}\sigma^{-2}_{n,\alpha}}$ \cite{Saga_2023}.

\section*{Setup}
To quantify the constraining power of the odd multipoles on the LPI-violating parameter $\alpha$, we consider the cross-correlations targeting ongoing and upcoming galaxy surveys, assuming maximal overlap of their survey regions.
Following Ref.~\cite{Saga_2023}, the surveys considered include the DESI, targeting Bright Galaxies (BGS), Luminous Red Galaxies (LRGs), and Emission Line Galaxies (ELGs) \cite{DESI2016}; Euclid, targeting H$\alpha$ emitters \cite{Euclid_2024}; the Subaru PFS, targeting [OII]-emitting ELGs \cite{PFS2014}; and the SKA, neutral atomic hydrogen (HI) galaxies in two planned phases, SKA1 and SKA2 \cite{Bacon_2020}. The number density and bias for cross-correlations between different survey samples in different redshift bins are estimated using the redshift slice of the sample with the larger bias as the reference, following Ref.~\cite{Saga_2022} (see Appendix E of that reference).
The survey volume of the overlapped regions is computed through $V=(4\pi/3)f_{\rm sky}\{r^3(z+\Delta z)-r^3(z-\Delta z)\}$, where $f_{\rm sky}$ is the fractional sky coverage, $r(z)$ is the comoving distance at redshift $z$, and $\Delta z$ is the redshift width of the smaller survey. In this work, we adopt the halo model framework, where each galaxy is assumed to reside in a dark matter halo. To estimate the halo potential $\phi_{\rm halo}$ and the galaxy velocity dispersion $v^2_{\rm g}$, we take the average over the mass range $[M_{\rm min},\infty]$, denoted by the tilde as
\begin{align}
\widetilde{f}(M_{\rm min},z,R_{\rm off}) = \frac{\int^{\infty}_{{\rm ln}M_{\rm min}}f(M,z,R_{\rm off})\frac{{\rm d}n}{{\rm dln}M}{\rm dln}M}{\int^{\infty}_{{\rm ln}M_{\rm min}}\frac{{\rm d}n}{{\rm dln}M}{\rm dln}M}
~. \label{eq:f_ave}
\end{align}
where $f=\{\phi_{\rm halo}, v^2_{\rm g}\}$, ${\rm d}n/{\rm dln} M$ represents the Sheth and Tormen mass function \cite{Sheth_1999}, and $M_{\rm min}$ denotes the minimum halo mass estimated from the galaxy bias \cite{Saga_2022}.

In the following, we set the side length of square pixels and maximum separation to $L_{\rm p}=2,{\rm Mpc}/h$ and $s_{\rm max}=30,{\rm Mpc}/h$, where the value of $s_{\rm max}$ corresponds to the scale where the gravitational redshift effect starts to dominate the signal.
We set the minimum separation, $s_{\rm min}$, to $s_{\rm min}\geq 5 {\rm Mpc}/h$ to neglect baryonic effects at small scales. 
Nonlinearity and non-Gaussian contributions could affect the constraints on $\alpha$ when setting $s_{\rm min}= 5 {\rm Mpc}/h$ as done in Ref.~\cite{Saga_2023}. Therefore, in our Fisher analysis, we also adopt $s_{\rm min}=15 {\rm Mpc}/h$ as a more conservative choice.

\begin{figure}
\includegraphics[width=\columnwidth]{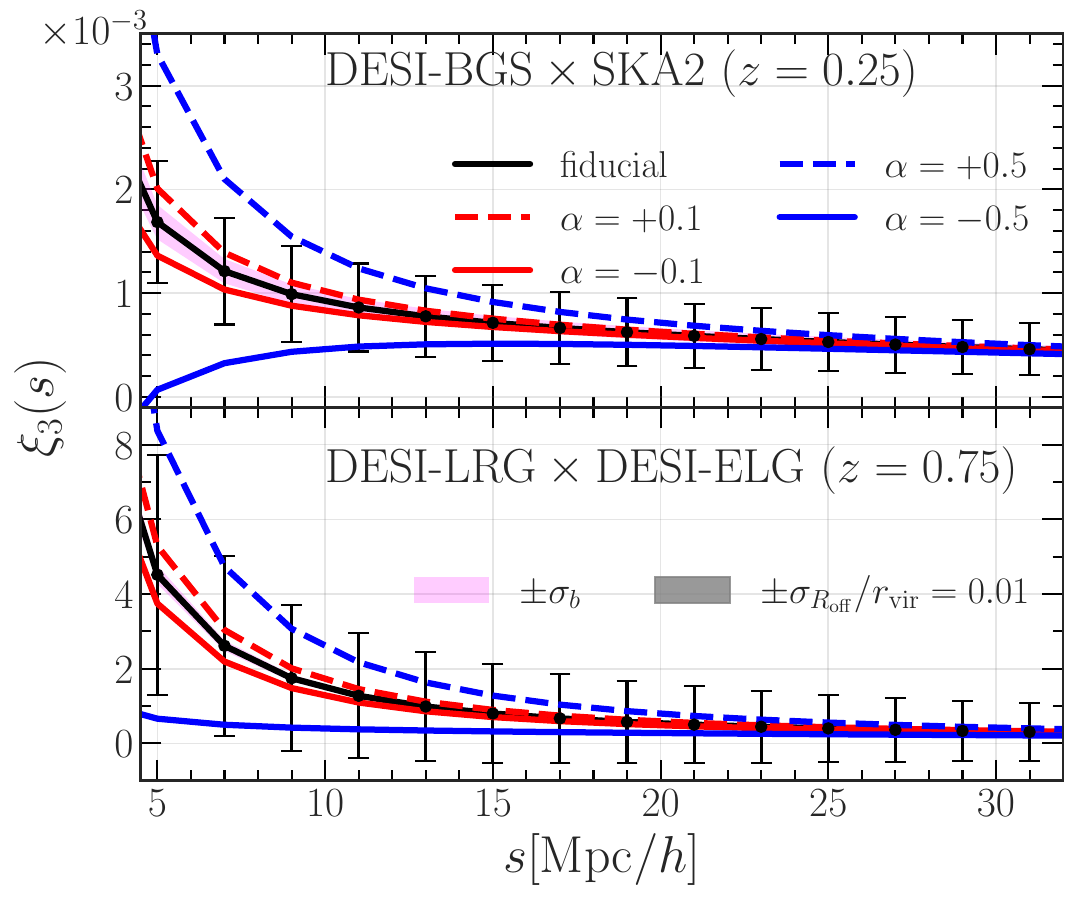}
\caption{Octupole moments along with their 1$\sigma$ errors expected from ${\rm DESI\mathchar`-BGS} \times {\rm SKA2}$ at $z=0.25$ (upper panel) and ${\rm DESI\mathchar`-LRG} \times {\rm DESI\mathchar`-ELG}$ at $z=0.75$ (lower panel).
The black lines represent the fiducial signal, while the blue and red lines correspond to $\alpha=\pm 0.5$ and $\alpha=\pm 0.1$, respectively. The gray and magenta shaded regions indicate the variation of the signal when varying $R_{\rm off}$ and $b$ within the expected 1$\sigma$ errors: $\sigma_{R_{\rm off}}/r_{\rm vir}=0.01$, $(\sigma_{b}^{\rm DESI\mathchar`-BGS}, \sigma_{b}^{\rm SKA2})=(0.05, 0.02)$, and $(\sigma_{b}^{\rm DESI\mathchar`-LRG}, \sigma_{b}^{\rm DESI\mathchar`-ELG})=(0.04, 0.02)$ \cite{Saga_2023}.}
\label{fig:xiggl}
\end{figure}
\section*{Results}
We present the forecasted constraints on the LPI-violating parameter $\alpha$, expected from the octupole alone and its combination with the dipole.
First, we demonstrate the constraining power of the octupole on $\alpha$, using the cross-correlation between different samples in a single redshift slice.
Figure~\ref{fig:xiggl} shows the octupole and its sensitivity to the parameters $\alpha$, $R_{\rm off}$, and $b$, as well as the expected $1\sigma$ errors.
The upper and lower panels present the results for ${\rm DESI\mathchar`-BGS} \times {\rm SKA2}$ at $z=0.25$ and ${\rm DESI\mathchar`-LRG} \times {\rm DESI\mathchar`-ELG}$ at $z=0.75$, respectively. Interestingly, we find that the octupole can constrain the LPI violation to $\alpha \lesssim 0.5$ for these ongoing and upcoming surveys when small-scale information (below $10 {\rm Mpc}/h$) is utilized, assuming the other parameters are fixed. Between the two analyses, ${\rm DESI\mathchar`-BGS} \times {\rm SKA2}$ provides tighter constraints than ${\rm DESI\mathchar`-LRG} \times {\rm DESI\mathchar`-ELG}$ because the BGS and SKA2 target galaxies with higher number densities, reducing the shot noise. 

\begin{figure}
\includegraphics[width=\columnwidth]{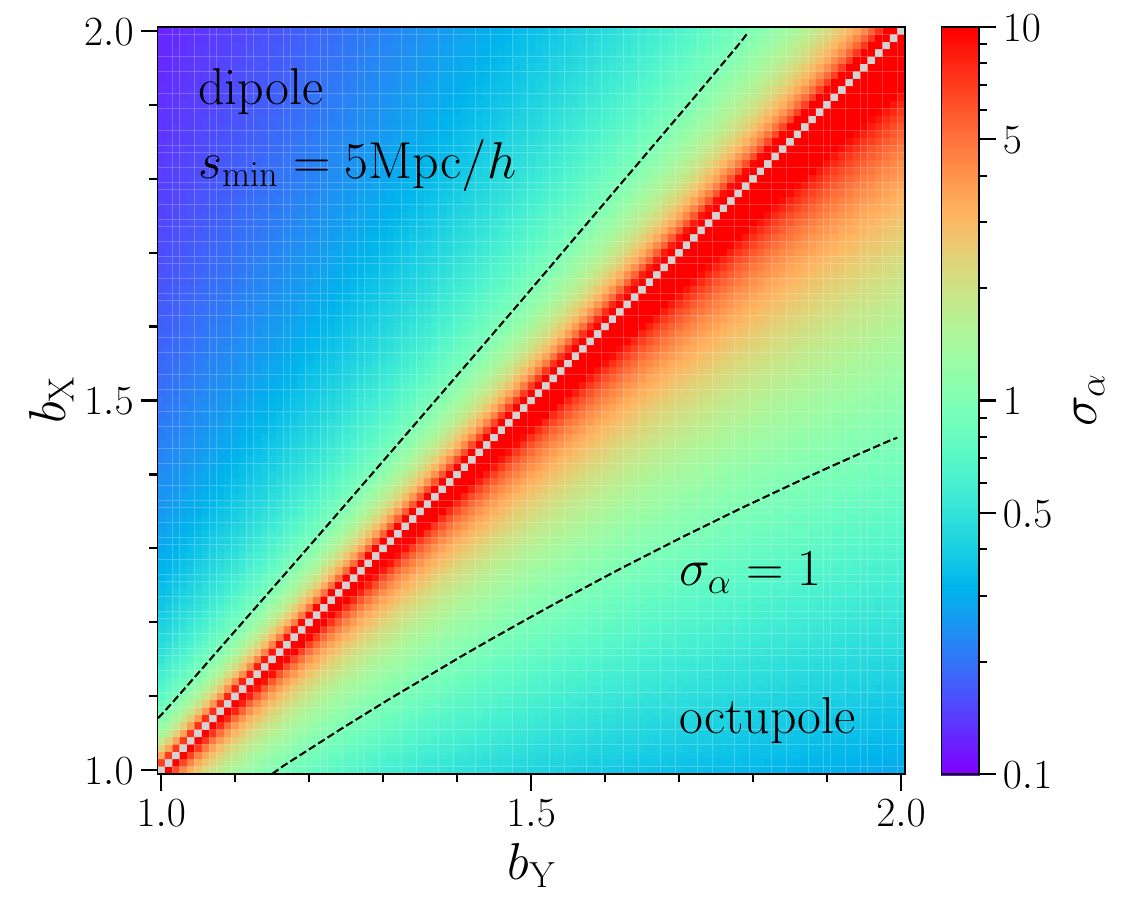}
\caption{$1 \sigma$ errors on the LPI-violating parameter $\alpha$ obtained from the odd multipoles at $z=0.5$ as a function of $b_{\rm X}$ and $b_{\rm Y}$. 
The regions with $b_{\rm X} > b_{\rm Y}$ and $b_{\rm Y} > b_{\rm X}$ correspond to the results obtained from the dipole and octupole, respectively. 
No constraint is obtained when $b_{\rm X} = b_{\rm Y}$.
The Gaussian priors $\sigma_{R_{\rm off}}$ and $\sigma_{b}$ are fixed to $\sigma_{R_{\rm off}}/r_{\rm vir}=\sigma_{b}/b=0.01$. $f_{\rm sky}$, $\Delta z$, and $s_{\rm min}$ are fixed to $f_{\rm sky}=1$, $\Delta z=0.1$, and $s_{\rm min}=5 {\rm Mpc}/h$, respectively.}
\label{fig:errorbxby}
\end{figure}

To see the sample dependence of the constraints in detail, we show the one-dimensional marginalized errors on $\alpha$ at $z=0.5$ as a function of two different bias parameters in Fig.~\ref{fig:errorbxby}. The results expected from the dipole and octupole are shown in the regions with $b_{\rm X} > b_{\rm Y}$ and $b_{\rm X} < b_{\rm Y}$, respectively.
We cannot constrain $\alpha$ if $b_{\rm X}=b_{\rm Y}$ where the odd multipoles vanish.
The halo mass and number density are estimated from the bias and redshift assuming the Sheth and Tormen mass function \cite{Sheth_1999}. Tighter constraints on $\alpha$ are obtained when cross-correlating samples with high and low biases, i.e., larger $\Delta b$ ($= b_{\rm X} - b_{\rm Y}$), for both odd multipole cases. 
Notably, the octupole alone can constrain the LPI violation to $\sigma_{\alpha} < 0.5$ in this setup, reflecting that the amplitude of the odd multipoles depends on the potential difference ($\Delta \phi_{\rm halo} = \phi_{\rm halo, X} - \phi_{\rm halo, Y}$), or equivalently the bias difference $\Delta b$ via the mass function, as shown in Eq.~(\ref{eq:ximu}). However, cross-correlating samples with high biases results in weaker constraints due to their lower number densities, which increase the shot noise contribution. 
Despite weaker signals, cross-correlations between samples with lower biases yield tighter constraints because of their smaller shot noise.
Furthermore, the constraints from the octupole of cross-correlations between low-bias samples become comparable to those from the dipole, increasing the benefit of combining them.
It is because the dipole and octupole have different bias dependencies. 
A term proportional to $b_{\rm X}b_{\rm Y}$, originating solely from the real-space contribution, is present in the dipole but absent in the octupole.
We set $s_{\rm min}=5 {\rm Mpc}/h$ here, but the choice of $s_{\rm min}$ does not affect the trend of the bias dependence, although the overall constraints on $\alpha$ change.
\begin{figure}
\includegraphics[width=\columnwidth]{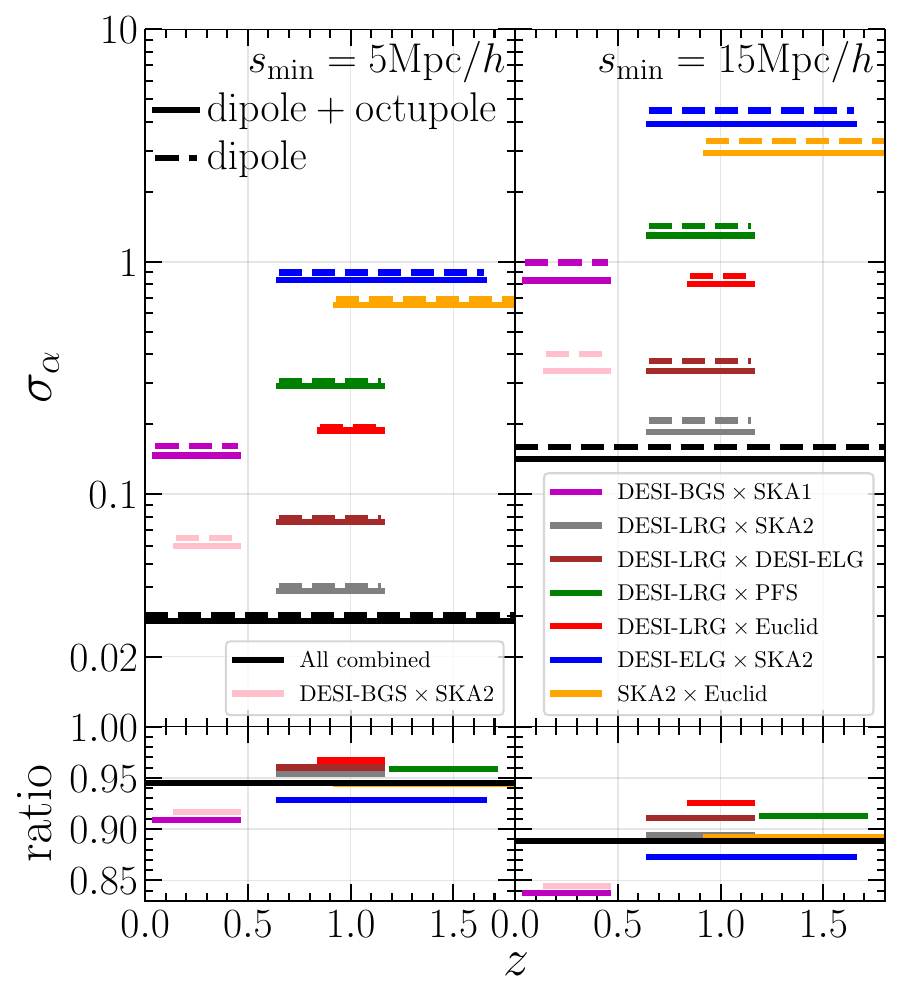}
\caption{{\it Top panels:} $1\sigma$ errors on the LPI-violating parameter $\alpha$ obtained from the dipole alone and its combination with the octupole, expected from the cross-correlations in the ongoing and upcoming surveys. 
The left and right panels show the results obtained from $s_{\rm min}=5 {\rm Mpc}/h$ and $15{\rm Mpc}/h$, respectively. {\it Bottom panels:} ratio of their constraints, $\sigma^{{\rm dipole+octupole}}_{\alpha}/\sigma^{{\rm dipole}}_{\alpha}$. The results of ${\rm DESI\mathchar`-LRG} \times {\rm PFS}$ are shifted horizontally for visibility.}
\label{fig:1derror}
\end{figure}

Finally, we show the benefit of utilizing the octupole to constrain $\alpha$ from ongoing and upcoming galaxy surveys, such as DESI, Subaru PFS, Euclid, and SKA1/2. 
The top panels of Fig.~\ref{fig:1derror} show the one-dimensional marginalized errors obtained from the dipole alone and its combination with the octupole, considering all the redshift contributions in the overlapping regions between the galaxy surveys. The bottom panels show the ratio of these results. The left and right panels show the results adopting $s_{\rm min}=5 {\rm Mpc}/h$ and $15{\rm Mpc}/h$, respectively. First, the results obtained from the dipole alone with $s_{\rm min}=5 {\rm Mpc}/h$ are consistent with those in Ref.~\cite{Saga_2023}. The tightest constraints are expected from the ${\rm DESI\mathchar`-LRG}$ and ${\rm SKA2}$ due to the large $\Delta b$, as well as the high number density in SKA2, for both $s_{\rm min}=5 {\rm Mpc}/h$ and $15{\rm Mpc}/h$. 
When all the constraints shown in Fig.~\ref{fig:1derror} are combined, 
we obtain the constraints on $\alpha$ using both the dipole and octupole as follows: $\sigma_{\alpha} \approx 0.028$ for $s_{\rm min} = 5 {\rm Mpc}/h$ and $\sigma_{\alpha} \approx 0.14$ for $s_{\rm min} = 15 {\rm Mpc}/h$. Next, looking at the ratio, we find that combining the dipole with the octupole improves the constraints on $\alpha$, with the benefit being pronounced when $s_{\rm min}=15 {\rm Mpc}/h$ for all the measurements considered in this analysis. 
The low-redshift surveys, namely ${\rm DESI\mathchar`-BGS}$ and ${\rm SKA1/2}$, particularly benefit from the octupole, via the cross-correlation between relatively low-bias samples, as discussed above. Remarkably, we obtain an average improvement of $6\%$ for $s_{\rm min}=5 {\rm Mpc}/h$ and $11\%$ for $s_{\rm min}=15 {\rm Mpc}/h$. 
We emphasize that these improvements offer notable gains without requiring additional data or more complex analysis.
Although the constraints on $\alpha$ for $s_{\rm min}=5 {\rm Mpc}/h$ are tighter than those for $s_{\rm min}=15 {\rm Mpc}/h$, the latter is the more conservative setup for our theoretical model neglecting the nonlinearity and non-Gaussian contributions for the density field. 

\section*{Conclusions}
High-precision data from ongoing and upcoming galaxy surveys offer a unique test of the equivalence principle on cosmological scales through the gravitational redshift effect in galaxy clustering. 
In this paper, focusing on galaxy clustering at small scales, we have explored the constraining power of the next leading odd multipole, the octupole, induced by the interplay between the Doppler and gravitational redshift effects for constraining the LPI-violating parameter $\alpha$.

Based on the Fisher analysis using the theoretical model of the cross-correlation function between two different samples, we have shown that not only is the octupole alone useful to constrain $\alpha$, but also its combination with the dipole further improves constraints for ongoing and upcoming surveys.
The improvements are particularly significant for cross-correlation between samples with low biases.
Although the constraints become tighter with the inclusion of small-scale information (i.e., small $s_{\rm min}$), we have demonstrated that the benefit of the octupole increases when restricting the analysis to larger scales, where our linear theory model is conservatively applied. 
For $s_{\rm min}=15 {\rm Mpc}/h$, ongoing and upcoming surveys are expected to improve by $11\%$ on average.
Our results indicate the significance of higher-order multipoles for testing the LPI on cosmological scales.

Finally, we comment on the precision of the octupole used in our analysis. While the theoretical model for the dipole has been well confirmed with simulation-based measurements~\cite{Saga_2022}, a verification of octupole prediction requires a more proper simulation setup.
In our future work, we plan to incorporate high-precision measurements based on $N$-body simulations to investigate the validity of the theoretical model.

\begin{acknowledgments} 
T.O. acknowledges the support of the Taiwan National Science and Technology Council under Grants No. NSTC NSTC 112-2112-M-001-034 and NSTC 113-2112-M-001-011, and the Academia Sinica Investigator Project Grant No. AS-IV-114-M03 for the period of 2025–2029. 
S.S. acknowledges support from the Japan Society for the Promotion of Science KAKENHI
Grants No. JP23K19050 and No. JP24K17043.
This work was supported in part by MEXT/JSPSKAKENHI Grants No. JP20H05861 and No. JP21H01081 (A.T.). 
\end{acknowledgments}

\bibliography{main_v2.bbl}

\providecommand{\noopsort}[1]{}\providecommand{\singleletter}[1]{#1}%
\begin{thebibliography}{69}
\expandafter\ifx\csname natexlab\endcsname\relax\def\natexlab#1{#1}\fi
\expandafter\ifx\csname bibnamefont\endcsname\relax
  \def\bibnamefont#1{#1}\fi
\expandafter\ifx\csname bibfnamefont\endcsname\relax
  \def\bibfnamefont#1{#1}\fi
\expandafter\ifx\csname citenamefont\endcsname\relax
  \def\citenamefont#1{#1}\fi
\expandafter\ifx\csname url\endcsname\relax
  \def\url#1{\texttt{#1}}\fi
\expandafter\ifx\csname urlprefix\endcsname\relax\def\urlprefix{URL }\fi
\providecommand{\bibinfo}[2]{#2}
\providecommand{\eprint}[2][]{\url{#2}}

\bibitem[{\citenamefont{{Sargent} and {Turner}}(1977)}]{Sargent_1977}
\bibinfo{author}{\bibfnamefont{W.~L.~W.} \bibnamefont{{Sargent}}} \bibnamefont{and} \bibinfo{author}{\bibfnamefont{E.~L.} \bibnamefont{{Turner}}}, \bibinfo{journal}{\apjl} \textbf{\bibinfo{volume}{212}}, \bibinfo{pages}{L3} (\bibinfo{year}{1977}).

\bibitem[{\citenamefont{{Peebles}}(1980)}]{Peebles_1980}
\bibinfo{author}{\bibfnamefont{P.~J.~E.} \bibnamefont{{Peebles}}}, \emph{\bibinfo{title}{{The large-scale structure of the universe}}} (\bibinfo{publisher}{Princeton University Press}, \bibinfo{year}{1980}).

\bibitem[{\citenamefont{{Kaiser}}(1987)}]{Kaiser_1987}
\bibinfo{author}{\bibfnamefont{N.}~\bibnamefont{{Kaiser}}}, \bibinfo{journal}{\mnras} \textbf{\bibinfo{volume}{227}}, \bibinfo{pages}{1} (\bibinfo{year}{1987}).

\bibitem[{\citenamefont{{Linder}}(2008)}]{Linder_2008}
\bibinfo{author}{\bibfnamefont{E.~V.} \bibnamefont{{Linder}}}, \bibinfo{journal}{Astroparticle Physics} \textbf{\bibinfo{volume}{29}}, \bibinfo{pages}{336} (\bibinfo{year}{2008}).

\bibitem[{\citenamefont{{Guzzo} et~al.}(2008)\citenamefont{{Guzzo}, {Pierleoni}, {Meneux}, {Branchini}, {Le F{\`e}vre}, {Marinoni}, {Garilli}, {Blaizot}, {De Lucia}, {Pollo} et~al.}}]{Guzzo_2008}
\bibinfo{author}{\bibfnamefont{L.}~\bibnamefont{{Guzzo}}}, \bibinfo{author}{\bibfnamefont{M.}~\bibnamefont{{Pierleoni}}}, \bibinfo{author}{\bibfnamefont{B.}~\bibnamefont{{Meneux}}}, \bibinfo{author}{\bibfnamefont{E.}~\bibnamefont{{Branchini}}}, \bibinfo{author}{\bibfnamefont{O.}~\bibnamefont{{Le F{\`e}vre}}}, \bibinfo{author}{\bibfnamefont{C.}~\bibnamefont{{Marinoni}}}, \bibinfo{author}{\bibfnamefont{B.}~\bibnamefont{{Garilli}}}, \bibinfo{author}{\bibfnamefont{J.}~\bibnamefont{{Blaizot}}}, \bibinfo{author}{\bibfnamefont{G.}~\bibnamefont{{De Lucia}}}, \bibinfo{author}{\bibfnamefont{A.}~\bibnamefont{{Pollo}}}, \bibnamefont{et~al.}, \bibinfo{journal}{\nat} \textbf{\bibinfo{volume}{451}}, \bibinfo{pages}{541} (\bibinfo{year}{2008}).

\bibitem[{\citenamefont{{Alam} et~al.}(2017{\natexlab{a}})\citenamefont{{Alam}, {Ata}, {Bailey}, {Beutler}, {Bizyaev}, {Blazek}, {Bolton}, {Brownstein}, {Burden}, {Chuang} et~al.}}]{Alam_2017}
\bibinfo{author}{\bibfnamefont{S.}~\bibnamefont{{Alam}}}, \bibinfo{author}{\bibfnamefont{M.}~\bibnamefont{{Ata}}}, \bibinfo{author}{\bibfnamefont{S.}~\bibnamefont{{Bailey}}}, \bibinfo{author}{\bibfnamefont{F.}~\bibnamefont{{Beutler}}}, \bibinfo{author}{\bibfnamefont{D.}~\bibnamefont{{Bizyaev}}}, \bibinfo{author}{\bibfnamefont{J.~A.} \bibnamefont{{Blazek}}}, \bibinfo{author}{\bibfnamefont{A.~S.} \bibnamefont{{Bolton}}}, \bibinfo{author}{\bibfnamefont{J.~R.} \bibnamefont{{Brownstein}}}, \bibinfo{author}{\bibfnamefont{A.}~\bibnamefont{{Burden}}}, \bibinfo{author}{\bibfnamefont{C.-H.} \bibnamefont{{Chuang}}}, \bibnamefont{et~al.}, \bibinfo{journal}{\mnras} \textbf{\bibinfo{volume}{470}}, \bibinfo{pages}{2617} (\bibinfo{year}{2017}{\natexlab{a}}).

\bibitem[{\citenamefont{{Sasaki}}(1987)}]{Sasaki_1987}
\bibinfo{author}{\bibfnamefont{M.}~\bibnamefont{{Sasaki}}}, \bibinfo{journal}{\mnras} \textbf{\bibinfo{volume}{228}}, \bibinfo{pages}{653} (\bibinfo{year}{1987}).

\bibitem[{\citenamefont{{Matsubara}}(2000)}]{Matsubara_2000}
\bibinfo{author}{\bibfnamefont{T.}~\bibnamefont{{Matsubara}}}, \bibinfo{journal}{\apjl} \textbf{\bibinfo{volume}{537}}, \bibinfo{pages}{L77} (\bibinfo{year}{2000}).

\bibitem[{\citenamefont{{Yoo} et~al.}(2009)\citenamefont{{Yoo}, {Fitzpatrick}, and {Zaldarriaga}}}]{Yoo_2009}
\bibinfo{author}{\bibfnamefont{J.}~\bibnamefont{{Yoo}}}, \bibinfo{author}{\bibfnamefont{A.~L.} \bibnamefont{{Fitzpatrick}}}, \bibnamefont{and} \bibinfo{author}{\bibfnamefont{M.}~\bibnamefont{{Zaldarriaga}}}, \bibinfo{journal}{\prd} \textbf{\bibinfo{volume}{80}}, \bibinfo{eid}{083514} (\bibinfo{year}{2009}).

\bibitem[{\citenamefont{{Bonvin} and {Durrer}}(2011)}]{Bonvin_2011}
\bibinfo{author}{\bibfnamefont{C.}~\bibnamefont{{Bonvin}}} \bibnamefont{and} \bibinfo{author}{\bibfnamefont{R.}~\bibnamefont{{Durrer}}}, \bibinfo{journal}{\prd} \textbf{\bibinfo{volume}{84}}, \bibinfo{eid}{063505} (\bibinfo{year}{2011}).

\bibitem[{\citenamefont{{Challinor} and {Lewis}}(2011)}]{Challinor_2011}
\bibinfo{author}{\bibfnamefont{A.}~\bibnamefont{{Challinor}}} \bibnamefont{and} \bibinfo{author}{\bibfnamefont{A.}~\bibnamefont{{Lewis}}}, \bibinfo{journal}{\prd} \textbf{\bibinfo{volume}{84}}, \bibinfo{eid}{043516} (\bibinfo{year}{2011}).

\bibitem[{\citenamefont{Yoo et~al.}(2012)\citenamefont{Yoo, Hamaus, Seljak, and Zaldarriaga}}]{Yoo_2012}
\bibinfo{author}{\bibfnamefont{J.}~\bibnamefont{Yoo}}, \bibinfo{author}{\bibfnamefont{N.}~\bibnamefont{Hamaus}}, \bibinfo{author}{\bibfnamefont{U.~c.~v.} \bibnamefont{Seljak}}, \bibnamefont{and} \bibinfo{author}{\bibfnamefont{M.}~\bibnamefont{Zaldarriaga}}, \bibinfo{journal}{Phys. Rev. D} \textbf{\bibinfo{volume}{86}}, \bibinfo{pages}{063514} (\bibinfo{year}{2012}).

\bibitem[{\citenamefont{{Jeong} et~al.}(2012)\citenamefont{{Jeong}, {Schmidt}, and {Hirata}}}]{Jeong_2012}
\bibinfo{author}{\bibfnamefont{D.}~\bibnamefont{{Jeong}}}, \bibinfo{author}{\bibfnamefont{F.}~\bibnamefont{{Schmidt}}}, \bibnamefont{and} \bibinfo{author}{\bibfnamefont{C.~M.} \bibnamefont{{Hirata}}}, \bibinfo{journal}{\prd} \textbf{\bibinfo{volume}{85}}, \bibinfo{eid}{023504} (\bibinfo{year}{2012}).

\bibitem[{\citenamefont{{Yoo}}(2014)}]{Yoo_2014}
\bibinfo{author}{\bibfnamefont{J.}~\bibnamefont{{Yoo}}}, \bibinfo{journal}{Classical and Quantum Gravity} \textbf{\bibinfo{volume}{31}}, \bibinfo{eid}{234001} (\bibinfo{year}{2014}).

\bibitem[{\citenamefont{{Raccanelli} et~al.}(2018)\citenamefont{{Raccanelli}, {Bertacca}, {Jeong}, {Neyrinck}, and {Szalay}}}]{Raccanelli_2016}
\bibinfo{author}{\bibfnamefont{A.}~\bibnamefont{{Raccanelli}}}, \bibinfo{author}{\bibfnamefont{D.}~\bibnamefont{{Bertacca}}}, \bibinfo{author}{\bibfnamefont{D.}~\bibnamefont{{Jeong}}}, \bibinfo{author}{\bibfnamefont{M.~C.} \bibnamefont{{Neyrinck}}}, \bibnamefont{and} \bibinfo{author}{\bibfnamefont{A.~S.} \bibnamefont{{Szalay}}}, \bibinfo{journal}{Physics of the Dark Universe} \textbf{\bibinfo{volume}{19}}, \bibinfo{pages}{109} (\bibinfo{year}{2018}).

\bibitem[{\citenamefont{{McDonald}}(2009)}]{McDonald_2009}
\bibinfo{author}{\bibfnamefont{P.}~\bibnamefont{{McDonald}}}, \bibinfo{journal}{\jcap} \textbf{\bibinfo{volume}{11}}, \bibinfo{eid}{026} (\bibinfo{year}{2009}).

\bibitem[{\citenamefont{{Croft}}(2013)}]{Croft_2013}
\bibinfo{author}{\bibfnamefont{R.~A.~C.} \bibnamefont{{Croft}}}, \bibinfo{journal}{\mnras} \textbf{\bibinfo{volume}{434}}, \bibinfo{pages}{3008} (\bibinfo{year}{2013}).

\bibitem[{\citenamefont{{Bonvin}}(2014)}]{Bonvin_2014_1}
\bibinfo{author}{\bibfnamefont{C.}~\bibnamefont{{Bonvin}}}, \bibinfo{journal}{Classical and Quantum Gravity} \textbf{\bibinfo{volume}{31}}, \bibinfo{eid}{234002} (\bibinfo{year}{2014}).

\bibitem[{\citenamefont{{Bonvin} et~al.}(2014)\citenamefont{{Bonvin}, {Hui}, and {Gazta{\~n}aga}}}]{Bonvin_2014}
\bibinfo{author}{\bibfnamefont{C.}~\bibnamefont{{Bonvin}}}, \bibinfo{author}{\bibfnamefont{L.}~\bibnamefont{{Hui}}}, \bibnamefont{and} \bibinfo{author}{\bibfnamefont{E.}~\bibnamefont{{Gazta{\~n}aga}}}, \bibinfo{journal}{\prd} \textbf{\bibinfo{volume}{89}}, \bibinfo{eid}{083535} (\bibinfo{year}{2014}).

\bibitem[{\citenamefont{{Kodwani} and {Desmond}}(2019)}]{Kodwani_2019}
\bibinfo{author}{\bibfnamefont{D.}~\bibnamefont{{Kodwani}}} \bibnamefont{and} \bibinfo{author}{\bibfnamefont{H.}~\bibnamefont{{Desmond}}}, \bibinfo{journal}{\prd} \textbf{\bibinfo{volume}{100}}, \bibinfo{eid}{064030} (\bibinfo{year}{2019}).

\bibitem[{\citenamefont{{Bonvin} and {Fleury}}(2018)}]{Bonvin_2018}
\bibinfo{author}{\bibfnamefont{C.}~\bibnamefont{{Bonvin}}} \bibnamefont{and} \bibinfo{author}{\bibfnamefont{P.}~\bibnamefont{{Fleury}}}, \bibinfo{journal}{\jcap} \textbf{\bibinfo{volume}{5}}, \bibinfo{eid}{061} (\bibinfo{year}{2018}).

\bibitem[{\citenamefont{{Bonvin} et~al.}(2020)\citenamefont{{Bonvin}, {Oliveira Franco}, and {Fleury}}}]{Bonvin_2020}
\bibinfo{author}{\bibfnamefont{C.}~\bibnamefont{{Bonvin}}}, \bibinfo{author}{\bibfnamefont{F.}~\bibnamefont{{Oliveira Franco}}}, \bibnamefont{and} \bibinfo{author}{\bibfnamefont{P.}~\bibnamefont{{Fleury}}}, \bibinfo{journal}{\jcap} \textbf{\bibinfo{volume}{2020}}, \bibinfo{eid}{004} (\bibinfo{year}{2020}).

\bibitem[{\citenamefont{{Umeh} et~al.}(2021)\citenamefont{{Umeh}, {Koyama}, and {Crittenden}}}]{Umeh_2021}
\bibinfo{author}{\bibfnamefont{O.}~\bibnamefont{{Umeh}}}, \bibinfo{author}{\bibfnamefont{K.}~\bibnamefont{{Koyama}}}, \bibnamefont{and} \bibinfo{author}{\bibfnamefont{R.}~\bibnamefont{{Crittenden}}}, \bibinfo{journal}{\jcap} \textbf{\bibinfo{volume}{2021}}, \bibinfo{eid}{049} (\bibinfo{year}{2021}).

\bibitem[{\citenamefont{{Wojtak} et~al.}(2011)\citenamefont{{Wojtak}, {Hansen}, and {Hjorth}}}]{Wojtak_2011}
\bibinfo{author}{\bibfnamefont{R.}~\bibnamefont{{Wojtak}}}, \bibinfo{author}{\bibfnamefont{S.~H.} \bibnamefont{{Hansen}}}, \bibnamefont{and} \bibinfo{author}{\bibfnamefont{J.}~\bibnamefont{{Hjorth}}}, \bibinfo{journal}{\nat} \textbf{\bibinfo{volume}{477}}, \bibinfo{pages}{567} (\bibinfo{year}{2011}).

\bibitem[{\citenamefont{{Sadeh} et~al.}(2015)\citenamefont{{Sadeh}, {Feng}, and {Lahav}}}]{Sadeh_2015}
\bibinfo{author}{\bibfnamefont{I.}~\bibnamefont{{Sadeh}}}, \bibinfo{author}{\bibfnamefont{L.~L.} \bibnamefont{{Feng}}}, \bibnamefont{and} \bibinfo{author}{\bibfnamefont{O.}~\bibnamefont{{Lahav}}}, \bibinfo{journal}{\prl} \textbf{\bibinfo{volume}{114}}, \bibinfo{eid}{071103} (\bibinfo{year}{2015}).

\bibitem[{\citenamefont{{Jimeno} et~al.}(2015)\citenamefont{{Jimeno}, {Broadhurst}, {Coupon}, {Umetsu}, and {Lazkoz}}}]{Jimeno_2015}
\bibinfo{author}{\bibfnamefont{P.}~\bibnamefont{{Jimeno}}}, \bibinfo{author}{\bibfnamefont{T.}~\bibnamefont{{Broadhurst}}}, \bibinfo{author}{\bibfnamefont{J.}~\bibnamefont{{Coupon}}}, \bibinfo{author}{\bibfnamefont{K.}~\bibnamefont{{Umetsu}}}, \bibnamefont{and} \bibinfo{author}{\bibfnamefont{R.}~\bibnamefont{{Lazkoz}}}, \bibinfo{journal}{\mnras} \textbf{\bibinfo{volume}{448}}, \bibinfo{pages}{1999} (\bibinfo{year}{2015}).

\bibitem[{\citenamefont{{Alam} et~al.}(2017{\natexlab{b}})\citenamefont{{Alam}, {Zhu}, {Croft}, {Ho}, {Giusarma}, and {Schneider}}}]{Alam_2017_gr}
\bibinfo{author}{\bibfnamefont{S.}~\bibnamefont{{Alam}}}, \bibinfo{author}{\bibfnamefont{H.}~\bibnamefont{{Zhu}}}, \bibinfo{author}{\bibfnamefont{R.~A.~C.} \bibnamefont{{Croft}}}, \bibinfo{author}{\bibfnamefont{S.}~\bibnamefont{{Ho}}}, \bibinfo{author}{\bibfnamefont{E.}~\bibnamefont{{Giusarma}}}, \bibnamefont{and} \bibinfo{author}{\bibfnamefont{D.~P.} \bibnamefont{{Schneider}}}, \bibinfo{journal}{\mnras} \textbf{\bibinfo{volume}{470}}, \bibinfo{pages}{2822} (\bibinfo{year}{2017}{\natexlab{b}}).

\bibitem[{\citenamefont{{Gaztanaga} et~al.}(2017)\citenamefont{{Gaztanaga}, {Bonvin}, and {Hui}}}]{Gaztanaga_2017}
\bibinfo{author}{\bibfnamefont{E.}~\bibnamefont{{Gaztanaga}}}, \bibinfo{author}{\bibfnamefont{C.}~\bibnamefont{{Bonvin}}}, \bibnamefont{and} \bibinfo{author}{\bibfnamefont{L.}~\bibnamefont{{Hui}}}, \bibinfo{journal}{\jcap} \textbf{\bibinfo{volume}{1}}, \bibinfo{eid}{032} (\bibinfo{year}{2017}).

\bibitem[{\citenamefont{{Zhu} et~al.}(2017)\citenamefont{{Zhu}, {Alam}, {Croft}, {Ho}, and {Giusarma}}}]{Zhu_2017}
\bibinfo{author}{\bibfnamefont{H.}~\bibnamefont{{Zhu}}}, \bibinfo{author}{\bibfnamefont{S.}~\bibnamefont{{Alam}}}, \bibinfo{author}{\bibfnamefont{R.~A.~C.} \bibnamefont{{Croft}}}, \bibinfo{author}{\bibfnamefont{S.}~\bibnamefont{{Ho}}}, \bibnamefont{and} \bibinfo{author}{\bibfnamefont{E.}~\bibnamefont{{Giusarma}}}, \bibinfo{journal}{\mnras} \textbf{\bibinfo{volume}{471}}, \bibinfo{pages}{2345} (\bibinfo{year}{2017}).

\bibitem[{\citenamefont{{Breton} et~al.}(2019)\citenamefont{{Breton}, {Rasera}, {Taruya}, {Lacombe}, and {Saga}}}]{Breton_2018}
\bibinfo{author}{\bibfnamefont{M.-A.} \bibnamefont{{Breton}}}, \bibinfo{author}{\bibfnamefont{Y.}~\bibnamefont{{Rasera}}}, \bibinfo{author}{\bibfnamefont{A.}~\bibnamefont{{Taruya}}}, \bibinfo{author}{\bibfnamefont{O.}~\bibnamefont{{Lacombe}}}, \bibnamefont{and} \bibinfo{author}{\bibfnamefont{S.}~\bibnamefont{{Saga}}}, \bibinfo{journal}{\mnras} \textbf{\bibinfo{volume}{483}}, \bibinfo{pages}{2671} (\bibinfo{year}{2019}).

\bibitem[{\citenamefont{{Beutler} and {Di Dio}}(2020)}]{Beutler_2020}
\bibinfo{author}{\bibfnamefont{F.}~\bibnamefont{{Beutler}}} \bibnamefont{and} \bibinfo{author}{\bibfnamefont{E.}~\bibnamefont{{Di Dio}}}, \bibinfo{journal}{\jcap} \textbf{\bibinfo{volume}{7}}, \bibinfo{eid}{048} (\bibinfo{year}{2020}).

\bibitem[{\citenamefont{{Saga} et~al.}(2020)\citenamefont{{Saga}, {Taruya}, {Breton}, and {Rasera}}}]{Saga_2020}
\bibinfo{author}{\bibfnamefont{S.}~\bibnamefont{{Saga}}}, \bibinfo{author}{\bibfnamefont{A.}~\bibnamefont{{Taruya}}}, \bibinfo{author}{\bibfnamefont{M.-A.} \bibnamefont{{Breton}}}, \bibnamefont{and} \bibinfo{author}{\bibfnamefont{Y.}~\bibnamefont{{Rasera}}}, \bibinfo{journal}{\mnras} \textbf{\bibinfo{volume}{498}}, \bibinfo{pages}{981} (\bibinfo{year}{2020}).

\bibitem[{\citenamefont{{Saga} et~al.}(2022)\citenamefont{{Saga}, {Taruya}, {Rasera}, and {Breton}}}]{Saga_2022}
\bibinfo{author}{\bibfnamefont{S.}~\bibnamefont{{Saga}}}, \bibinfo{author}{\bibfnamefont{A.}~\bibnamefont{{Taruya}}}, \bibinfo{author}{\bibfnamefont{Y.}~\bibnamefont{{Rasera}}}, \bibnamefont{and} \bibinfo{author}{\bibfnamefont{M.-A.} \bibnamefont{{Breton}}}, \bibinfo{journal}{\mnras} \textbf{\bibinfo{volume}{511}}, \bibinfo{pages}{2732} (\bibinfo{year}{2022}).

\bibitem[{\citenamefont{{Giusarma} et~al.}(2017)\citenamefont{{Giusarma}, {Alam}, {Zhu}, {Croft}, and {Ho}}}]{Giusarma_2017}
\bibinfo{author}{\bibfnamefont{E.}~\bibnamefont{{Giusarma}}}, \bibinfo{author}{\bibfnamefont{S.}~\bibnamefont{{Alam}}}, \bibinfo{author}{\bibfnamefont{H.}~\bibnamefont{{Zhu}}}, \bibinfo{author}{\bibfnamefont{R.~A.~C.} \bibnamefont{{Croft}}}, \bibnamefont{and} \bibinfo{author}{\bibfnamefont{S.}~\bibnamefont{{Ho}}}, \bibinfo{journal} \bibinfo{eid}{arXiv:1709.07854} (\bibinfo{year}{2017}).

\bibitem[{\citenamefont{{Di Dio} and {Seljak}}(2019)}]{Dio_2019}
\bibinfo{author}{\bibfnamefont{E.}~\bibnamefont{{Di Dio}}} \bibnamefont{and} \bibinfo{author}{\bibfnamefont{U.}~\bibnamefont{{Seljak}}}, \bibinfo{journal}{\jcap} \textbf{\bibinfo{volume}{4}}, \bibinfo{eid}{050} (\bibinfo{year}{2019}).

\bibitem[{\citenamefont{{Takada} et~al.}(2014)\citenamefont{{Takada}, {Ellis}, {Chiba}, {Greene}, {Aihara}, {Arimoto}, {Bundy}, {Cohen}, {Dor{\'e}}, {Graves} et~al.}}]{PFS2014}
\bibinfo{author}{\bibfnamefont{M.}~\bibnamefont{{Takada}}}, \bibinfo{author}{\bibfnamefont{R.~S.} \bibnamefont{{Ellis}}}, \bibinfo{author}{\bibfnamefont{M.}~\bibnamefont{{Chiba}}}, \bibinfo{author}{\bibfnamefont{J.~E.} \bibnamefont{{Greene}}}, \bibinfo{author}{\bibfnamefont{H.}~\bibnamefont{{Aihara}}}, \bibinfo{author}{\bibfnamefont{N.}~\bibnamefont{{Arimoto}}}, \bibinfo{author}{\bibfnamefont{K.}~\bibnamefont{{Bundy}}}, \bibinfo{author}{\bibfnamefont{J.}~\bibnamefont{{Cohen}}}, \bibinfo{author}{\bibfnamefont{O.}~\bibnamefont{{Dor{\'e}}}}, \bibinfo{author}{\bibfnamefont{G.}~\bibnamefont{{Graves}}}, \bibnamefont{et~al.}, \bibinfo{journal}{\pasj} \textbf{\bibinfo{volume}{66}}, \bibinfo{eid}{R1} (\bibinfo{year}{2014}).

\bibitem[{\citenamefont{{DESI Collaboration} et~al.}(2016)\citenamefont{{DESI Collaboration}, {Aghamousa}, {Aguilar}, {Ahlen}, {Alam}, {Allen}, {Allende Prieto}, {Annis}, {Bailey}, {Balland} et~al.}}]{DESI2016}
\bibinfo{author}{\bibnamefont{{DESI Collaboration}}}, \bibinfo{author}{\bibfnamefont{A.}~\bibnamefont{{Aghamousa}}}, \bibinfo{author}{\bibfnamefont{J.}~\bibnamefont{{Aguilar}}}, \bibinfo{author}{\bibfnamefont{S.}~\bibnamefont{{Ahlen}}}, \bibinfo{author}{\bibfnamefont{S.}~\bibnamefont{{Alam}}}, \bibinfo{author}{\bibfnamefont{L.~E.} \bibnamefont{{Allen}}}, \bibinfo{author}{\bibfnamefont{C.}~\bibnamefont{{Allende Prieto}}}, \bibinfo{author}{\bibfnamefont{J.}~\bibnamefont{{Annis}}}, \bibinfo{author}{\bibfnamefont{S.}~\bibnamefont{{Bailey}}}, \bibinfo{author}{\bibfnamefont{C.}~\bibnamefont{{Balland}}}, \bibnamefont{et~al.}, \bibinfo{journal} \bibinfo{eid}{arXiv:1611.00036} (\bibinfo{year}{2016}).

\bibitem[{\citenamefont{{Euclid Collaboration} et~al.}(2024)\citenamefont{{Euclid Collaboration}, {Mellier}, {Abdurro'uf}, {Acevedo Barroso}, {Ach{\'u}carro}, {Adamek}, {Adam}, {Addison}, {Aghanim}, {Aguena} et~al.}}]{Euclid_2024}
\bibinfo{author}{\bibnamefont{{Euclid Collaboration}}}, \bibinfo{author}{\bibfnamefont{Y.}~\bibnamefont{{Mellier}}}, \bibinfo{author}{\bibnamefont{{Abdurro'uf}}}, \bibinfo{author}{\bibfnamefont{J.~A.} \bibnamefont{{Acevedo Barroso}}}, \bibinfo{author}{\bibfnamefont{A.}~\bibnamefont{{Ach{\'u}carro}}}, \bibinfo{author}{\bibfnamefont{J.}~\bibnamefont{{Adamek}}}, \bibinfo{author}{\bibfnamefont{R.}~\bibnamefont{{Adam}}}, \bibinfo{author}{\bibfnamefont{G.~E.} \bibnamefont{{Addison}}}, \bibinfo{author}{\bibfnamefont{N.}~\bibnamefont{{Aghanim}}}, \bibinfo{author}{\bibfnamefont{M.}~\bibnamefont{{Aguena}}}, \bibnamefont{et~al.}, \bibinfo{journal} \bibinfo{eid}{arXiv:2405.13491} (\bibinfo{year}{2024}).

\bibitem[{\citenamefont{{Square Kilometre Array Cosmology Science Working Group} et~al.}(2020)\citenamefont{{Square Kilometre Array Cosmology Science Working Group}, {Bacon}, {Battye}, {Bull}, {Camera}, {Ferreira}, {Harrison}, {Parkinson}, {Pourtsidou}, {Santos} et~al.}}]{Bacon_2020}
\bibinfo{author}{\bibnamefont{{Square Kilometre Array Cosmology Science Working Group}}}, \bibinfo{author}{\bibfnamefont{D.~J.} \bibnamefont{{Bacon}}}, \bibinfo{author}{\bibfnamefont{R.~A.} \bibnamefont{{Battye}}}, \bibinfo{author}{\bibfnamefont{P.}~\bibnamefont{{Bull}}}, \bibinfo{author}{\bibfnamefont{S.}~\bibnamefont{{Camera}}}, \bibinfo{author}{\bibfnamefont{P.~G.} \bibnamefont{{Ferreira}}}, \bibinfo{author}{\bibfnamefont{I.}~\bibnamefont{{Harrison}}}, \bibinfo{author}{\bibfnamefont{D.}~\bibnamefont{{Parkinson}}}, \bibinfo{author}{\bibfnamefont{A.}~\bibnamefont{{Pourtsidou}}}, \bibinfo{author}{\bibfnamefont{M.~G.} \bibnamefont{{Santos}}}, \bibnamefont{et~al.}, \bibinfo{journal}{Publications of the Astronomical Society of Australia} \textbf{\bibinfo{volume}{37}}, \bibinfo{eid}{e007} (\bibinfo{year}{2020}).

\bibitem[{\citenamefont{{Bonvin} et~al.}(2023)\citenamefont{{Bonvin}, {Lepori}, {Schulz}, {Tutusaus}, {Adamek}, and {Fosalba}}}]{Bonvin_2023}
\bibinfo{author}{\bibfnamefont{C.}~\bibnamefont{{Bonvin}}}, \bibinfo{author}{\bibfnamefont{F.}~\bibnamefont{{Lepori}}}, \bibinfo{author}{\bibfnamefont{S.}~\bibnamefont{{Schulz}}}, \bibinfo{author}{\bibfnamefont{I.}~\bibnamefont{{Tutusaus}}}, \bibinfo{author}{\bibfnamefont{J.}~\bibnamefont{{Adamek}}}, \bibnamefont{and} \bibinfo{author}{\bibfnamefont{P.}~\bibnamefont{{Fosalba}}}, \bibinfo{journal}{\mnras} \textbf{\bibinfo{volume}{525}}, \bibinfo{pages}{4611} (\bibinfo{year}{2023}).

\bibitem[{\citenamefont{{Lepori} et~al.}(2024)\citenamefont{{Lepori}, {Schulz}, {Tutusaus}, {Breton}, {Saga}, {Viglione}, {Adamek}, {Bonvin}, {Dam}, {Fosalba} et~al.}}]{Lepori_2024}
\bibinfo{author}{\bibfnamefont{F.}~\bibnamefont{{Lepori}}}, \bibinfo{author}{\bibfnamefont{S.}~\bibnamefont{{Schulz}}}, \bibinfo{author}{\bibfnamefont{I.}~\bibnamefont{{Tutusaus}}}, \bibinfo{author}{\bibfnamefont{M.~A.} \bibnamefont{{Breton}}}, \bibinfo{author}{\bibfnamefont{S.}~\bibnamefont{{Saga}}}, \bibinfo{author}{\bibfnamefont{C.}~\bibnamefont{{Viglione}}}, \bibinfo{author}{\bibfnamefont{J.}~\bibnamefont{{Adamek}}}, \bibinfo{author}{\bibfnamefont{C.}~\bibnamefont{{Bonvin}}}, \bibinfo{author}{\bibfnamefont{L.}~\bibnamefont{{Dam}}}, \bibinfo{author}{\bibfnamefont{P.}~\bibnamefont{{Fosalba}}}, \bibnamefont{et~al.}, \bibinfo{journal} \bibinfo{eid}{arXiv:2410.06268} (\bibinfo{year}{2024}).

\bibitem[{\citenamefont{{Saga} et~al.}(2023)\citenamefont{{Saga}, {Taruya}, {Rasera}, and {Breton}}}]{Saga_2023}
\bibinfo{author}{\bibfnamefont{S.}~\bibnamefont{{Saga}}}, \bibinfo{author}{\bibfnamefont{A.}~\bibnamefont{{Taruya}}}, \bibinfo{author}{\bibfnamefont{Y.}~\bibnamefont{{Rasera}}}, \bibnamefont{and} \bibinfo{author}{\bibfnamefont{M.-A.} \bibnamefont{{Breton}}}, \bibinfo{journal}{\mnras} \textbf{\bibinfo{volume}{524}}, \bibinfo{pages}{4472} (\bibinfo{year}{2023}).

\bibitem[{\citenamefont{{Pound} and {Rebka}}(1959)}]{Pound_1959}
\bibinfo{author}{\bibfnamefont{R.~V.} \bibnamefont{{Pound}}} \bibnamefont{and} \bibinfo{author}{\bibfnamefont{G.~A.} \bibnamefont{{Rebka}}}, \bibinfo{journal}{\prl} \textbf{\bibinfo{volume}{3}}, \bibinfo{pages}{439} (\bibinfo{year}{1959}).

\bibitem[{\citenamefont{{Pound} and {Snider}}(1965)}]{Pound_1965}
\bibinfo{author}{\bibfnamefont{R.~V.} \bibnamefont{{Pound}}} \bibnamefont{and} \bibinfo{author}{\bibfnamefont{J.~L.} \bibnamefont{{Snider}}}, \bibinfo{journal}{Physical Review} \textbf{\bibinfo{volume}{140}}, \bibinfo{pages}{B788} (\bibinfo{year}{1965}).

\bibitem[{\citenamefont{Will}(2018)}]{Will_2018}
\bibinfo{author}{\bibfnamefont{C.~M.} \bibnamefont{Will}}, \emph{\bibinfo{title}{Theory and Experiment in Gravitational Physics}} (\bibinfo{publisher}{Cambridge University Press}, \bibinfo{year}{2018}), \bibinfo{edition}{2nd} ed.

\bibitem[{\citenamefont{{Bekenstein}}(1982)}]{Bekenstein_1982}
\bibinfo{author}{\bibfnamefont{J.~D.} \bibnamefont{{Bekenstein}}}, \bibinfo{journal}{\prd} \textbf{\bibinfo{volume}{25}}, \bibinfo{pages}{1527} (\bibinfo{year}{1982}).

\bibitem[{\citenamefont{{Sandvik} et~al.}(2002)\citenamefont{{Sandvik}, {Barrow}, and {Magueijo}}}]{Sandvik_2002}
\bibinfo{author}{\bibfnamefont{H.~B.} \bibnamefont{{Sandvik}}}, \bibinfo{author}{\bibfnamefont{J.~D.} \bibnamefont{{Barrow}}}, \bibnamefont{and} \bibinfo{author}{\bibfnamefont{J.}~\bibnamefont{{Magueijo}}}, \bibinfo{journal}{\prl} \textbf{\bibinfo{volume}{88}}, \bibinfo{eid}{031302} (\bibinfo{year}{2002}).

\bibitem[{\citenamefont{{Barrow} and {Magueijo}}(1998)}]{Barrow_1998}
\bibinfo{author}{\bibfnamefont{J.~D.} \bibnamefont{{Barrow}}} \bibnamefont{and} \bibinfo{author}{\bibfnamefont{J.}~\bibnamefont{{Magueijo}}}, \bibinfo{journal}{Physics Letters B} \textbf{\bibinfo{volume}{443}}, \bibinfo{pages}{104} (\bibinfo{year}{1998}).

\bibitem[{\citenamefont{{Barrow}}(1999)}]{Barrow_1999}
\bibinfo{author}{\bibfnamefont{J.~D.} \bibnamefont{{Barrow}}}, \bibinfo{journal}{\prd} \textbf{\bibinfo{volume}{59}}, \bibinfo{eid}{043515} (\bibinfo{year}{1999}).

\bibitem[{\citenamefont{{Blanco} et~al.}(2024)\citenamefont{{Blanco}, {Bonvin}, {Clarkson}, and {Maartens}}}]{Blanco_2024}
\bibinfo{author}{\bibfnamefont{D.~S.} \bibnamefont{{Blanco}}}, \bibinfo{author}{\bibfnamefont{C.}~\bibnamefont{{Bonvin}}}, \bibinfo{author}{\bibfnamefont{C.}~\bibnamefont{{Clarkson}}}, \bibnamefont{and} \bibinfo{author}{\bibfnamefont{R.}~\bibnamefont{{Maartens}}}, \bibinfo{journal}{\jcap} \textbf{\bibinfo{volume}{12}}, \bibinfo{eid}{029} (\bibinfo{year}{2024}).

\bibitem[{\citenamefont{{Komatsu} et~al.}(2011)\citenamefont{{Komatsu}, {Smith}, {Dunkley}, {Bennett}, {Gold}, {Hinshaw}, {Jarosik}, {Larson}, {Nolta}, {Page} et~al.}}]{Komatsu:2011}
\bibinfo{author}{\bibfnamefont{E.}~\bibnamefont{{Komatsu}}}, \bibinfo{author}{\bibfnamefont{K.~M.} \bibnamefont{{Smith}}}, \bibinfo{author}{\bibfnamefont{J.}~\bibnamefont{{Dunkley}}}, \bibinfo{author}{\bibfnamefont{C.~L.} \bibnamefont{{Bennett}}}, \bibinfo{author}{\bibfnamefont{B.}~\bibnamefont{{Gold}}}, \bibinfo{author}{\bibfnamefont{G.}~\bibnamefont{{Hinshaw}}}, \bibinfo{author}{\bibfnamefont{N.}~\bibnamefont{{Jarosik}}}, \bibinfo{author}{\bibfnamefont{D.}~\bibnamefont{{Larson}}}, \bibinfo{author}{\bibfnamefont{M.~R.} \bibnamefont{{Nolta}}}, \bibinfo{author}{\bibfnamefont{L.}~\bibnamefont{{Page}}}, \bibnamefont{et~al.}, \bibinfo{journal}{\apjs} \textbf{\bibinfo{volume}{192}}, \bibinfo{eid}{18} (\bibinfo{year}{2011}).

\bibitem[{\citenamefont{{Zhao} et~al.}(2013)\citenamefont{{Zhao}, {Peacock}, and {Li}}}]{Zhao_2013}
\bibinfo{author}{\bibfnamefont{H.}~\bibnamefont{{Zhao}}}, \bibinfo{author}{\bibfnamefont{J.~A.} \bibnamefont{{Peacock}}}, \bibnamefont{and} \bibinfo{author}{\bibfnamefont{B.}~\bibnamefont{{Li}}}, \bibinfo{journal}{\prd} \textbf{\bibinfo{volume}{88}}, \bibinfo{eid}{043013} (\bibinfo{year}{2013}).

\bibitem[{\citenamefont{{Kaiser}}(2013)}]{Kaiser_2013}
\bibinfo{author}{\bibfnamefont{N.}~\bibnamefont{{Kaiser}}}, \bibinfo{journal}{\mnras} \textbf{\bibinfo{volume}{435}}, \bibinfo{pages}{1278} (\bibinfo{year}{2013}).

\bibitem[{\citenamefont{{Cai} et~al.}(2017)\citenamefont{{Cai}, {Kaiser}, {Cole}, and {Frenk}}}]{Cai_2017}
\bibinfo{author}{\bibfnamefont{Y.-C.} \bibnamefont{{Cai}}}, \bibinfo{author}{\bibfnamefont{N.}~\bibnamefont{{Kaiser}}}, \bibinfo{author}{\bibfnamefont{S.}~\bibnamefont{{Cole}}}, \bibnamefont{and} \bibinfo{author}{\bibfnamefont{C.}~\bibnamefont{{Frenk}}}, \bibinfo{journal}{\mnras} \textbf{\bibinfo{volume}{468}}, \bibinfo{pages}{1981} (\bibinfo{year}{2017}).

\bibitem[{\citenamefont{{Navarro} et~al.}(1996)\citenamefont{{Navarro}, {Frenk}, and {White}}}]{Navarro_1996}
\bibinfo{author}{\bibfnamefont{J.~F.} \bibnamefont{{Navarro}}}, \bibinfo{author}{\bibfnamefont{C.~S.} \bibnamefont{{Frenk}}}, \bibnamefont{and} \bibinfo{author}{\bibfnamefont{S.~D.~M.} \bibnamefont{{White}}}, \bibinfo{journal}{\apj} \textbf{\bibinfo{volume}{462}}, \bibinfo{pages}{563} (\bibinfo{year}{1996}).

\bibitem[{\citenamefont{{{\L}okas} and {Mamon}}(2001)}]{Lokas_2001}
\bibinfo{author}{\bibfnamefont{E.~L.} \bibnamefont{{{\L}okas}}} \bibnamefont{and} \bibinfo{author}{\bibfnamefont{G.~A.} \bibnamefont{{Mamon}}}, \bibinfo{journal}{\mnras} \textbf{\bibinfo{volume}{321}}, \bibinfo{pages}{155} (\bibinfo{year}{2001}).

\bibitem[{\citenamefont{{Sheth} and {Diaferio}}(2001)}]{Sheth_2001}
\bibinfo{author}{\bibfnamefont{R.~K.} \bibnamefont{{Sheth}}} \bibnamefont{and} \bibinfo{author}{\bibfnamefont{A.}~\bibnamefont{{Diaferio}}}, \bibinfo{journal}{\mnras} \textbf{\bibinfo{volume}{322}}, \bibinfo{pages}{901} (\bibinfo{year}{2001}).

\bibitem[{\citenamefont{{Hikage} et~al.}(2013)\citenamefont{{Hikage}, {Mandelbaum}, {Takada}, and {Spergel}}}]{Hikage_2013}
\bibinfo{author}{\bibfnamefont{C.}~\bibnamefont{{Hikage}}}, \bibinfo{author}{\bibfnamefont{R.}~\bibnamefont{{Mandelbaum}}}, \bibinfo{author}{\bibfnamefont{M.}~\bibnamefont{{Takada}}}, \bibnamefont{and} \bibinfo{author}{\bibfnamefont{D.~N.} \bibnamefont{{Spergel}}}, \bibinfo{journal}{\mnras} \textbf{\bibinfo{volume}{435}}, \bibinfo{pages}{2345} (\bibinfo{year}{2013}).

\bibitem[{\citenamefont{{Masaki} et~al.}(2013)\citenamefont{{Masaki}, {Hikage}, {Takada}, {Spergel}, and {Sugiyama}}}]{Masaki_2013}
\bibinfo{author}{\bibfnamefont{S.}~\bibnamefont{{Masaki}}}, \bibinfo{author}{\bibfnamefont{C.}~\bibnamefont{{Hikage}}}, \bibinfo{author}{\bibfnamefont{M.}~\bibnamefont{{Takada}}}, \bibinfo{author}{\bibfnamefont{D.~N.} \bibnamefont{{Spergel}}}, \bibnamefont{and} \bibinfo{author}{\bibfnamefont{N.}~\bibnamefont{{Sugiyama}}}, \bibinfo{journal}{\mnras} \textbf{\bibinfo{volume}{433}}, \bibinfo{pages}{3506} (\bibinfo{year}{2013}).

\bibitem[{\citenamefont{{Yan} et~al.}(2020)\citenamefont{{Yan}, {Raza}, {Van Waerbeke}, {Mead}, {McCarthy}, {Tr{\"o}ster}, and {Hinshaw}}}]{Yan_2020}
\bibinfo{author}{\bibfnamefont{Z.}~\bibnamefont{{Yan}}}, \bibinfo{author}{\bibfnamefont{N.}~\bibnamefont{{Raza}}}, \bibinfo{author}{\bibfnamefont{L.}~\bibnamefont{{Van Waerbeke}}}, \bibinfo{author}{\bibfnamefont{A.~J.} \bibnamefont{{Mead}}}, \bibinfo{author}{\bibfnamefont{I.~G.} \bibnamefont{{McCarthy}}}, \bibinfo{author}{\bibfnamefont{T.}~\bibnamefont{{Tr{\"o}ster}}}, \bibnamefont{and} \bibinfo{author}{\bibfnamefont{G.}~\bibnamefont{{Hinshaw}}}, \bibinfo{journal}{\mnras} \textbf{\bibinfo{volume}{493}}, \bibinfo{pages}{1120} (\bibinfo{year}{2020}).

\bibitem[{\citenamefont{{Sheth} and {Tormen}}(1999)}]{Sheth_1999}
\bibinfo{author}{\bibfnamefont{R.~K.} \bibnamefont{{Sheth}}} \bibnamefont{and} \bibinfo{author}{\bibfnamefont{G.}~\bibnamefont{{Tormen}}}, \bibinfo{journal}{\mnras} \textbf{\bibinfo{volume}{308}}, \bibinfo{pages}{119} (\bibinfo{year}{1999}).

\bibitem[{\citenamefont{{Taruya} et~al.}(2020)\citenamefont{{Taruya}, {Saga}, {Breton}, {Rasera}, and {Fujita}}}]{Taruya_2019}
\bibinfo{author}{\bibfnamefont{A.}~\bibnamefont{{Taruya}}}, \bibinfo{author}{\bibfnamefont{S.}~\bibnamefont{{Saga}}}, \bibinfo{author}{\bibfnamefont{M.-A.} \bibnamefont{{Breton}}}, \bibinfo{author}{\bibfnamefont{Y.}~\bibnamefont{{Rasera}}}, \bibnamefont{and} \bibinfo{author}{\bibfnamefont{T.}~\bibnamefont{{Fujita}}}, \bibinfo{journal}{\mnras} \textbf{\bibinfo{volume}{491}}, \bibinfo{pages}{4162} (\bibinfo{year}{2020}).

\bibitem[{\citenamefont{{Reimberg} et~al.}(2016)\citenamefont{{Reimberg}, {Bernardeau}, and {Pitrou}}}]{Reimberg_2016}
\bibinfo{author}{\bibfnamefont{P.}~\bibnamefont{{Reimberg}}}, \bibinfo{author}{\bibfnamefont{F.}~\bibnamefont{{Bernardeau}}}, \bibnamefont{and} \bibinfo{author}{\bibfnamefont{C.}~\bibnamefont{{Pitrou}}}, \bibinfo{journal}{\jcap} \textbf{\bibinfo{volume}{1}}, \bibinfo{pages}{048} (\bibinfo{year}{2016}).

\bibitem[{\citenamefont{{Castorina} and {White}}(2018{\natexlab{a}})}]{Castorina_2018a}
\bibinfo{author}{\bibfnamefont{E.}~\bibnamefont{{Castorina}}} \bibnamefont{and} \bibinfo{author}{\bibfnamefont{M.}~\bibnamefont{{White}}}, \bibinfo{journal}{\mnras} \textbf{\bibinfo{volume}{476}}, \bibinfo{pages}{4403} (\bibinfo{year}{2018}{\natexlab{a}}).

\bibitem[{\citenamefont{{Castorina} and {White}}(2018{\natexlab{b}})}]{Castorina_2018b}
\bibinfo{author}{\bibfnamefont{E.}~\bibnamefont{{Castorina}}} \bibnamefont{and} \bibinfo{author}{\bibfnamefont{M.}~\bibnamefont{{White}}}, \bibinfo{journal}{\mnras} \textbf{\bibinfo{volume}{479}}, \bibinfo{pages}{741} (\bibinfo{year}{2018}{\natexlab{b}}).

\bibitem[{\citenamefont{{Tegmark}}(1997)}]{Tegmark_1997}
\bibinfo{author}{\bibfnamefont{M.}~\bibnamefont{{Tegmark}}}, \bibinfo{journal}{\prl} \textbf{\bibinfo{volume}{79}}, \bibinfo{pages}{3806} (\bibinfo{year}{1997}).

\bibitem[{\citenamefont{{Tegmark} et~al.}(1997)\citenamefont{{Tegmark}, {Taylor}, and {Heavens}}}]{Tegmark_etal_1997}
\bibinfo{author}{\bibfnamefont{M.}~\bibnamefont{{Tegmark}}}, \bibinfo{author}{\bibfnamefont{A.~N.} \bibnamefont{{Taylor}}}, \bibnamefont{and} \bibinfo{author}{\bibfnamefont{A.~F.} \bibnamefont{{Heavens}}}, \bibinfo{journal}{\apj} \textbf{\bibinfo{volume}{480}}, \bibinfo{pages}{22} (\bibinfo{year}{1997}).

\bibitem[{\citenamefont{{Hall} and {Bonvin}}(2017)}]{Hall_2017}
\bibinfo{author}{\bibfnamefont{A.}~\bibnamefont{{Hall}}} \bibnamefont{and} \bibinfo{author}{\bibfnamefont{C.}~\bibnamefont{{Bonvin}}}, \bibinfo{journal}{\prd} \textbf{\bibinfo{volume}{95}}, \bibinfo{eid}{043530} (\bibinfo{year}{2017}).

\bibitem[{\citenamefont{{Lepori} et~al.}(2018)\citenamefont{{Lepori}, {Di Dio}, {Villa}, and {Viel}}}]{Lepori_2018}
\bibinfo{author}{\bibfnamefont{F.}~\bibnamefont{{Lepori}}}, \bibinfo{author}{\bibfnamefont{E.}~\bibnamefont{{Di Dio}}}, \bibinfo{author}{\bibfnamefont{E.}~\bibnamefont{{Villa}}}, \bibnamefont{and} \bibinfo{author}{\bibfnamefont{M.}~\bibnamefont{{Viel}}}, \bibinfo{journal}{\jcap} \textbf{\bibinfo{volume}{5}}, \bibinfo{eid}{043} (\bibinfo{year}{2018}).

\end{thebibliography}

\end{document}